\documentclass[aps,10pt,superscriptaddress,twocolumn,nofootinbib]{revtex4-2}

\usepackage{siunitx}
\usepackage{array,amsmath,amsthm,feynmf,mathtools}
\usepackage{multirow}
\usepackage{mathtools}
\usepackage{slashed}
\usepackage{amssymb}
\usepackage{bm}
\usepackage{cancel}
\usepackage[normalem]{ulem}
\usepackage[export]{adjustbox}
\usepackage{enumitem}
\usepackage{float}

\usepackage{graphicx}
\usepackage{xcolor}
\definecolor{CiteBlue}{RGB}{45,52,151}
\usepackage[
    colorlinks=true,
    linkcolor=CiteBlue,
    urlcolor=CiteBlue,
    citecolor=CiteBlue
]{hyperref}

\usepackage[capitalise]{cleveref}

\allowdisplaybreaks

\begin{document}

\title{Probing New Physics with $\mu^+ \mu^- \to bs$ at a Muon Collider}
\author{Wolfgang Altmannshofer}
\email{waltmann@ucsc.edu}
\affiliation{Department of Physics and Santa Cruz Institute for Particle Physics\\ University of California, Santa Cruz, CA 95064, USA}

\author{Sri Aditya Gadam}
\email{sgadam@ucsc.edu}
\affiliation{Department of Physics and Santa Cruz Institute for Particle Physics\\ University of California, Santa Cruz, CA 95064, USA}

\author{Stefano Profumo}
\email{profumo@ucsc.edu}
\affiliation{Department of Physics and Santa Cruz Institute for Particle Physics\\ University of California, Santa Cruz, CA 95064, USA}

\begin{abstract}
We show that bottom-strange production at a high-energy muon collider, $\mu^+ \mu^- \to b s$, is a sensitive probe of new physics. We consider the full set of four-fermion contact interactions that contribute to this process at dimension 6, and discuss the complementarity of a muon collider and of the study of rare $B$ meson decays that also probe said new physics. 
If a signal were to be found at a muon collider, the forward-backward asymmetry of the $b$-jet provides diagnostics about the underlying chirality structure of the new physics. 
In the absence of a signal at a center of mass energy of $10$~TeV, $\mu^+ \mu^- \to b s$ can indirectly probe new physics at scales close to $100$~TeV. We also discuss the impact that beam polarization has on the muon collider sensitivity performance.
\end{abstract}

\maketitle

\section{Introduction} \label{sec:intro}

Rare decays of bottom quark ($b$) hadrons are widely acknowledged as important probes of Beyond the Standard Model (BSM) physics~\cite{Blake:2016olu, Altmannshofer:2022hfs}. Several experimental results on rare $b$ decays show poor agreement with the corresponding Standard Model (SM) predictions. The LHCb collaboration reports deviations in the angular distribution of the $B\to K^*\mu^+\mu^-$ decay (the ``$P_5^\prime$ anomaly'') and in the branching ratios of the $B_s \to \phi \mu^+\mu^-$,
$B\to K^*\mu^+\mu^-$, and $B\to K \mu^+\mu^-$ decays. Until recently, these hints for new physics were supported by LHCb results on $R_K$ and $R_{K^*}$ that showed evidence for lepton flavor universality violation in rare $B$ meson decays.
Global fits of rare $b$ decay data had found a remarkably consistent explanation of these ``$B$ anomalies'' in terms of new physics~\cite{Geng:2021nhg, Altmannshofer:2021qrr, Alguero:2021anc, Hurth:2021nsi, Ciuchini:2021smi, Gubernari:2022hxn}. 

However, the latest update from LHCb on $R_K$ and $R_{K^*}$ is in excellent agreement with the SM predictions~\cite{LHCb:2022qnv, LHCb:2022zom}, implying that the origin of the remaining $b \to s \mu \mu$ anomalies, be it SM or BSM physics, is to a good approximation universal for muons and electrons. If the anomalies are due to new physics, global fits point to the lepton flavor universal 4-fermion contact interaction $(\bar s \gamma_\alpha P_L b)(\bar \ell \gamma^\alpha \ell)$. As is well known, new physics in this contact interaction can, in principle, be mimicked by non-perturbative QCD effects. While recent calculations indicate that hadronic effects are under control~\cite{Gubernari:2020eft, Gubernari:2022hxn}, a SM origin of the anomalies cannot be excluded. In this context, it is highly motivated to consider additional probes of this type of 4-fermion contact interaction.

The $B$ anomalies hint at a new physics scale $\Lambda_\text{NP} \sim 35$~TeV for $\mathcal O(1)$ couplings. It is thus conceivable that if new physics is responsible for the rare $B$ decay anomalies, it is beyond the direct reach of current colliders. The description in terms of contact interactions might remain valid up to energies of 10's of TeV. Even in that case, there are model-independent signatures that can be predicted at colliders. 

For example, at proton-proton colliders, one can access the parton level $b s \to \mu^+\mu^- $ process and expect enhanced di-muon production at large di-muon invariant mass~\cite{Greljo:2017vvb}. However, the expected sensitivity at the high-luminosity LHC will be insufficient to test a heavy new physics origin of the rare $B$ anomalies in a model-independent fashion.

In a preliminary study, published as a Snowmass whitepaper~\cite{Altmannshofer:2022xri}, we showed that non-standard $\mu^+ \mu^- \to bs$ production could be observed with high significance at a 10 TeV muon collider if the $B$ anomalies are due to heavy new physics. Furthermore, the forward-backward asymmetry of the b-jet provides diagnostics of the chirality structure of the new physics couplings. The high sensitivity of the muon collider stems from the fact that the non-standard $\mu^+ \mu^- \to bs$ cross section increases with the center of mass energy, while the relevant background cross sections (mainly misidentified di-jets) decrease. For recent related studies see e.g.~\cite{Huang:2021nkl, Huang:2021biu, Asadi:2021gah, Qian:2021ihf, Azatov:2022itm, Sun:2023cuf}.

In this paper, we extend the analysis of~\cite{Altmannshofer:2022xri} and provide a detailed discussion of the prospects to probe new physics using $\mu^+ \mu^- \to bs$ at a muon collider.
In section~\ref{sec:framework}, we introduce the theoretical framework and outline the operators we include in our analysis. In contrast to~\cite{Altmannshofer:2022xri}, we include the full set of 4-fermion contact interactions that can contribute to $\mu^+ \mu^- \to bs$ at leading order.
In section~\ref{sec:status}, we review the status of the global rare $b$ decay fit after the recent LHCb update of $R_K$ and $R_{K^*}$. We identify two relevant scenarios: (i) new physics is lepton flavor universal and {it} addresses the remaining $b \to s \mu \mu$ anomalies; (ii) new physics is muon-specific, and thus strongly constrained. In both cases, we also discuss the expected sensitivity of rare $b$ decays after the high-luminosity phase of the LHC.
In section~\ref{sec:mumubs}, we discuss the differential $\mu^+ \mu^- \to bs$ cross-section, taking into account the full set of 4-fermion contact interactions and including the effect of muon beam polarization.  
In section~\ref{sec:collider}, we discuss all relevant sources of backgrounds, including the irreducible SM background as well as di-jet production where one of the jets is mis-tagged. Finally, in section~\ref{sec:SentivityProjections}, we provide the sensitivity projections for a future muon collider. Assuming the presence of new physics, we discuss how precisely a muon collider could identify a lepton-universal new physics effect. In the absence of new physics, we estimate the constraining power of a muon collider.
We also compare the muon collider sensitivity to the sensitivity from rare $b$ decays.
We conclude in section~\ref{sec:conclusions}. In appendices~\ref{sec:RGEs} and~\ref{sec:GlobalFit}, we give details about the renormalization group running of the Wilson coefficients and our global rare $b$ decay fit.

\section{Theoretical Framework}\label{sec:framework}

We employ the standard effective Hamiltonian framework that parametrizes the BSM contributions to the $b\rightarrow s \ell \ell$ decays at the scale of the $B$ mesons, considering all relevant dimension 6 operators along with their Wilson coefficients:
\begin{multline}
\label{eq:effectiveHamiltonian}
    \mathcal H_\text{eff} =  \mathcal H_\text{eff}^\text{SM} - \frac{4G_F}{\sqrt{2}} V_{tb} V_{ts}^* \frac{e^2}{16\pi^2} \\
    \times \Big( \Delta C_9^\ell O_9^\ell + \Delta C_{10}^\ell O_{10}^\ell + C^{\prime \ell}_9 O^{\prime \ell}_9 + C^{\prime \ell}_{10} O^{\prime \ell}_{10} \\
    + C_{S}^\ell O_S^\ell + C_P^\ell O_P^\ell
    + C^{\prime \ell}_{S} O^{\prime \ell}_S + C_P^{\prime \ell} O^{\prime \ell}_P \Big) ~,
\end{multline}
where $\ell$ runs over the three lepton flavors $e, \mu, \tau$.
This effective Hamiltonian retains the standard normalization factors, including the CKM factors that are typical of the SM contributions to this process. The operators can be expressed as 4-fermion contact structures and are
\begin{eqnarray}
O_9^\ell  &=&  (\overline{s} \gamma^\alpha P_L b) (\overline{\ell} \gamma_\alpha \ell) ~, \\
O_{10}^\ell  &=&  (\overline{s} \gamma^\alpha P_L b) (\overline{\ell} \gamma_\alpha \gamma_5 \ell) ~, \\
O_S^\ell &=& (\overline{s} P_R b) (\overline{\ell} \ell) ~, \\
O_P^\ell &=& (\overline{s} P_R b) (\overline{\ell} \gamma_5 \ell) ~.
\end{eqnarray}
The remaining operators, $O'$, can be obtained from the operators $O$ with the interchange $\{L \leftrightarrow R \}$. 

The SM Hamiltonian, $\mathcal H_\text{eff}^\text{SM}$, contains the operators $O_9^\ell$ and $O_{10}^\ell$ with lepton flavor universal Wilson coefficients $C_9^\text{SM} \simeq 4.2$ and $C_{10}^\text{SM} \simeq -4.1$. The corresponding new physics Wilson coefficients can in principle be lepton flavor specific and we denote them as $\Delta C_9^\ell$ and $\Delta C_{10}^\ell$. All the other operators are negligible in the SM. 

We assume that the new physics that sources the effective Hamiltonian in~\eqref{eq:effectiveHamiltonian} is sufficiently heavy compared to the center of mass energy of a muon collider, such that the effective framework remains applicable. We consider this a conservative assumption: If the new physics is lighter, it can be produced on-shell at the muon collider and it would be generically easier to detect~\cite{Huang:2021nkl, Huang:2021biu, Asadi:2021gah, Qian:2021ihf, Azatov:2022itm}.

At the high center of mass energy of a muon collider, the $SU(2)_L \times U(1)_Y$ gauge symmetry of the SM needs to be taken into account. This can be done by using the Standard Model Effective Field Theory (SMEFT)~\cite{Grzadkowski:2010es} to parameterize new physics at energies above the electroweak scale. As there are no $bs\ell\ell$ 4-fermion operators with tensor structures in SMEFT, we omitted them in the effective Hamiltonian~\eqref{eq:effectiveHamiltonian} to be consistent. Moreover, the scalar and pseudoscalar operators are related in SMEFT such that~\cite{Alonso:2014csa}
\begin{equation}
C_S^\ell = - C_P^\ell ~,\qquad C_S^{\prime \ell} = C_P^{\prime \ell} ~. 
\end{equation}
We will impose these relations throughout. 
One can expect generic corrections to these relations of order $v^2/\Lambda_\text{NP}^2$, where $v \simeq 246$\,GeV is the vacuum expectation value of the Higgs and $\Lambda_\text{NP}$ the new physics scale. As we will see, for $\mathcal O(1)$ new physics couplings, this ratio is of $\mathcal O(10^{-5})$, and in that case, the corrections are completely negligible.

The remaining operators in the effective Hamiltonian~\eqref{eq:effectiveHamiltonian} are unconstrained in SMEFT and the corresponding Wilson coefficients can be treated as independent parameters. Instead of using SMEFT notation, we prefer to adopt the low-energy notation of~\eqref{eq:effectiveHamiltonian} to facilitate the comparison between the muon collider and rare $B$ decays. 

For a precise sensitivity comparison, one should take into account the renormalization group running between the center of mass energy of a muon collider, $\mu \sim \sqrt{s}$, and the relevant energy scale for $B$ meson decays, typically chosen to be the $b$ mass, $\mu \sim m_b$. While these scales differ by more than 3 orders of magnitude, the impact of running is modest and usually does not exceed 10\%. Details are given in appendix~\ref{sec:RGEs}. 

\section{Status and Prospects of Rare B Decay Fits}\label{sec:status}

\begin{figure}[tb]
\centering
\includegraphics[width=1.0\linewidth]{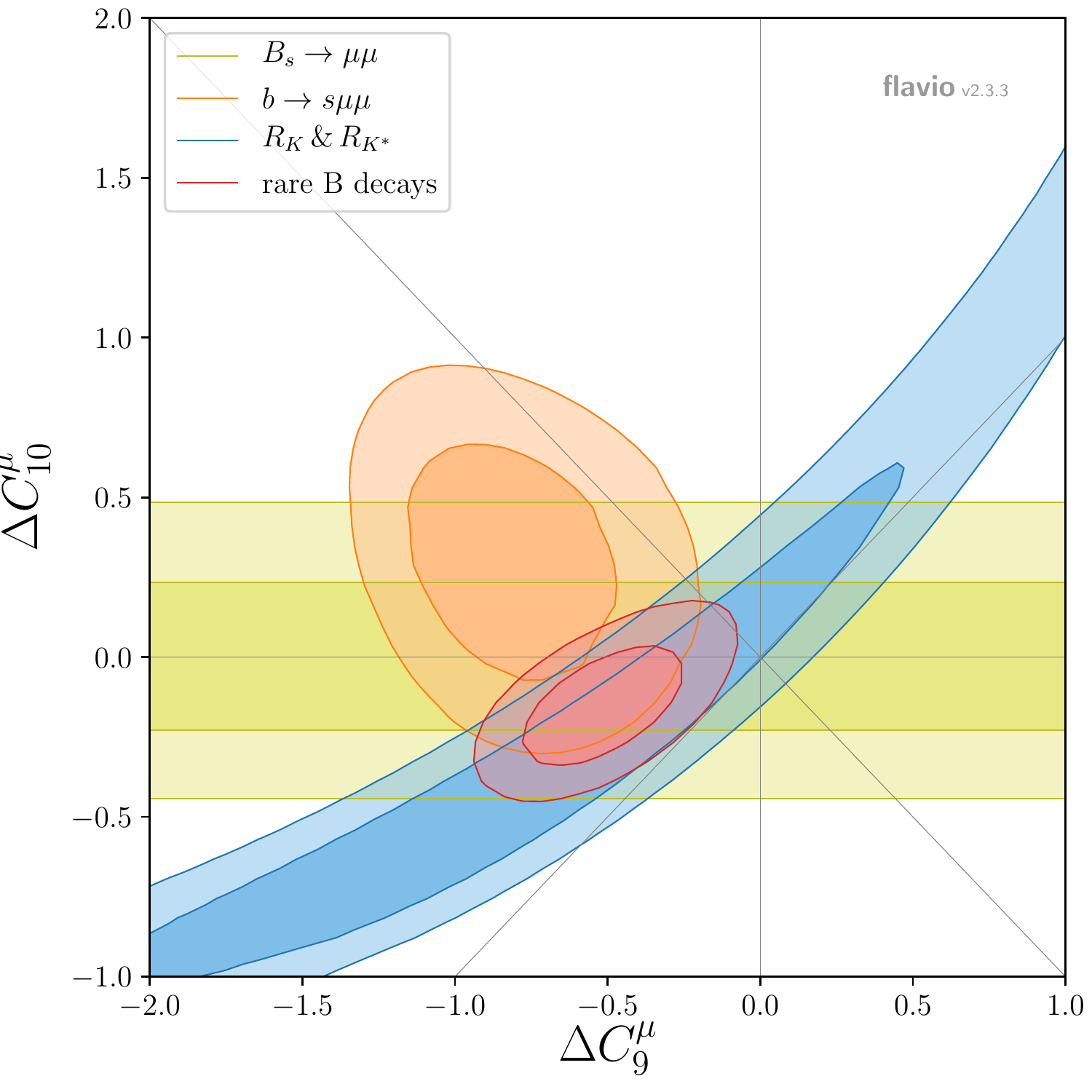}
\caption{The global rare $B$ decay fit in the plane of muon specific new physics contributions to $C_9$ and $C_{10}$, after the recent updates of $B_s \to \mu^+ \mu^-$~\cite{CMS:2022dbz} and $R_K$, $R_{K^*}$~\cite{LHCb:2022qnv, LHCb:2022zom}. The fit includes the $B_s \to \mu^+ \mu^-$ branching ratio (yellow), $R_K$, $R_{K^*}$ and other LFU tests (blue), and $B \to K \mu^+ \mu^-$, $B \to K^* \mu^+ \mu^-$, $B_s \to \phi \mu^+ \mu^-$, $\Lambda_b \to \Lambda \mu^+ \mu^-$ branching ratios and angular observables (orange). The result of the global fit is shown in red.}
\label{fig:global_fit}
\end{figure}

At a muon collider, we are interested in 4-fermion operators with $\ell = \mu$, while $B$ meson decays can, in principle, be used to probe all lepton flavors $\ell = e, \mu, \tau$. 
Until recently, rare $B$ decays have provided intriguing hints for new physics contributions to the muon-specific Wilson coefficients $\Delta C_9^\mu$ and $\Delta C_{10}^\mu$ (see e.g.~\cite{Geng:2021nhg, Altmannshofer:2021qrr, Alguero:2021anc, Hurth:2021nsi, Ciuchini:2021smi}). Those hints were based on a number of experimental results, in particular the anomalously low rates of the $B \to K \mu \mu$, $B \to K^* \mu \mu$, and $B_s \to \phi \mu \mu$  decays~\cite{LHCb:2014cxe, LHCb:2016ykl, LHCb:2021zwz}, the anomalous angular distribution of $B \to K^* \mu \mu$~\cite{LHCb:2020lmf, LHCb:2020gog} and the hints for lepton universality violation~\cite{LHCb:2017avl, LHCb:2021trn}.
However, the most recent results by LHCb on the lepton flavor universality ratios $R_K$ and $R_{K^*}$~\cite{LHCb:2022qnv, LHCb:2022zom} are in excellent agreement with the SM predictions and strongly constrain new physics in muon-specific Wilson coefficients. 

\begin{figure}[tb]
\centering
\includegraphics[width=1.0\linewidth]{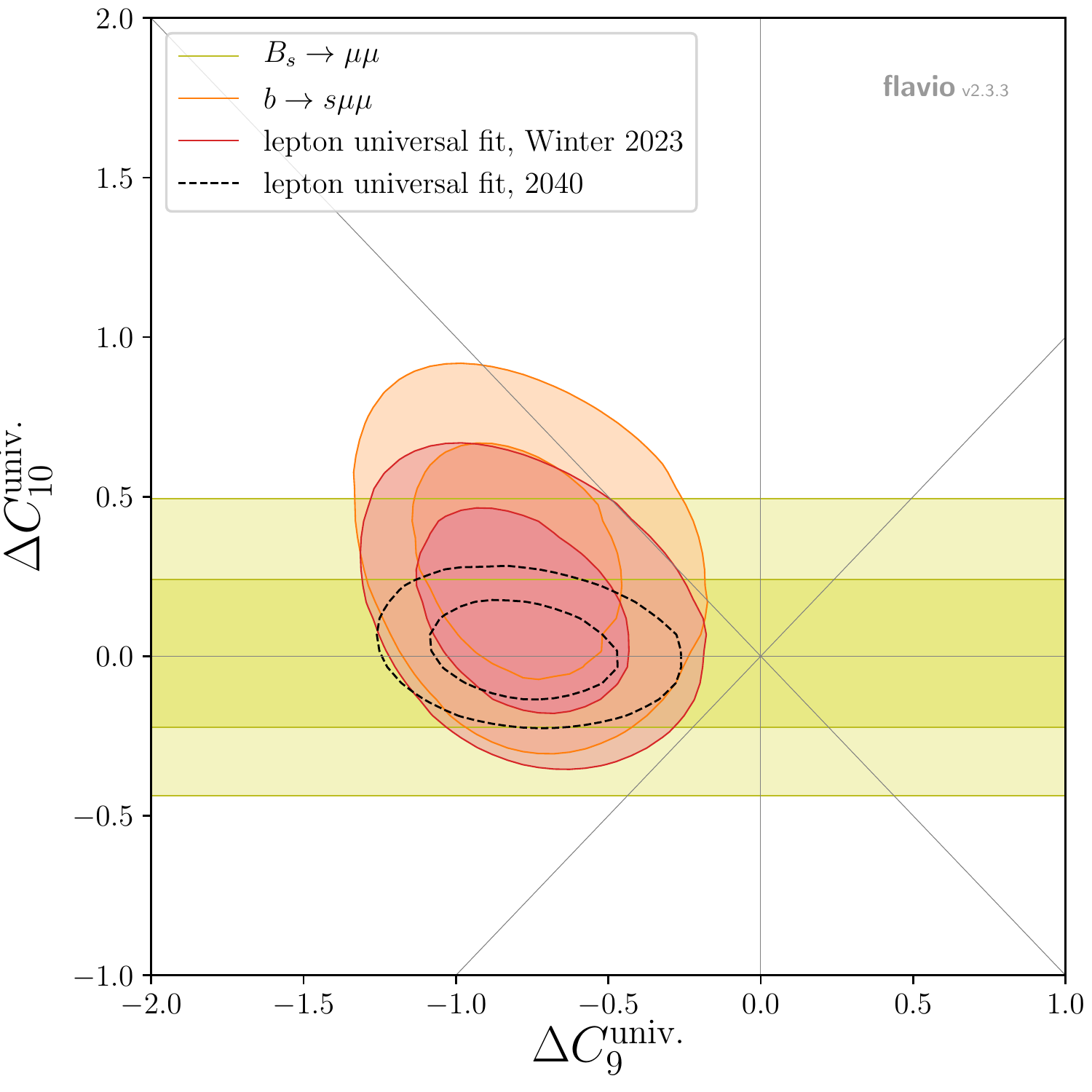} \\
\includegraphics[width=1.0\linewidth]{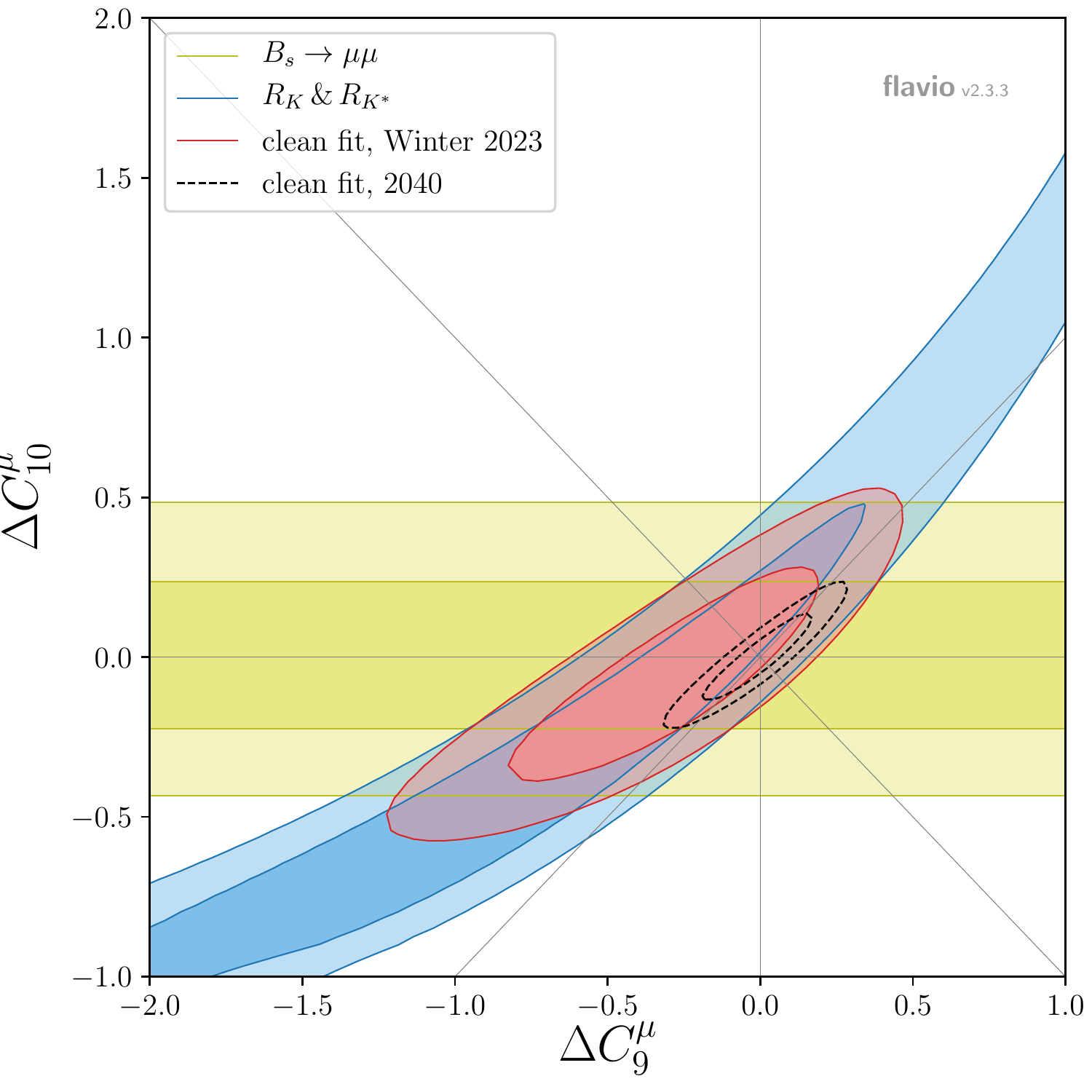}
\caption{Top: Rare $B$ decay fit in the plane of lepton universal new physics contributions to $C_9$ and $C_{10}$. Bottom: Rare $B$ decay fit in the plane of muon-specific new physics contributions to $C_9$ and $C_{10}$, including only theoretically clean observables.}
\label{fig:more_fits}
\end{figure}

The situation is summarized in figure~\ref{fig:global_fit} which shows the various constraints on the muon specific $\Delta C_9^\mu$ and $\Delta C_{10}^\mu$ (see also~\cite{Ciuchini:2022wbq, Greljo:2022jac, Allanach:2022iod, Alguero:2023jeh, Wen:2023pfq, Allanach:2023uxz} for related studies). Details on how this figure was obtained are given in appendix~\ref{sec:GlobalFit}. 
As in~\cite{Altmannshofer:2021qrr}, the observables are grouped into three categories: lepton flavor universality tests (blue), the $B_s \to \mu^+ \mu^-$ branching ratio (yellow), and the semileptonic $b \to s \mu \mu$ branching ratios and angular observables (orange). The combination is shown in red. Compared to the situation two years ago~\cite{Altmannshofer:2021qrr}, there is a tension among the different categories. Both the $B_s \to \mu^+ \mu^-$ branching ratio and the lepton flavor universality tests $R_K$ and $R_{K^*}$ are compatible with the SM predictions ($\Delta C_9^\mu = \Delta C_{10}^\mu = 0$), while the $b \to s \mu \mu$ observables do show a preference for a non-standard $\Delta C_9^\mu$.  

The tension in the fit can be resolved in two ways: (i) assuming that the new physics effect is lepton flavor universal; (ii) assuming that the SM predictions for the $b \to s \mu \mu$ observables are affected by unexpectedly large hadronic effects, rendering the corresponding region unreliable. 

In case (i), the constraints from the LFU tests, $R_K$ and $R_{K^*}$ in particular, do not apply, and one is left with the overlap of the $B_s \to \mu^+ \mu^-$ region and the $b \to s \mu^+\mu^-$ region. This situation is shown in the upper plot of figure~\ref{fig:more_fits}. Approximating the likelihood in the vicinity of the best-fit point by a multivariate Gaussian, we find
\begin{equation} \label{eq:C9C10_benchmark}
  \Delta C_9^\text{univ.} = -0.81 \pm 0.22 ~,\quad \Delta C_{10}^\text{univ.} = +0.12 \pm 0.20~,
\end{equation}
with an error correlation of $\rho = -30\%$.
This corresponds to a $\sim 2.8\sigma$ preference for new physics, mainly in $\Delta C_9^\text{univ.}$. Such a lepton-universal shift in $C_9$ can, in principle, be mimicked by a hadronic effect in the rare $B$ decays. Therefore it is highly motivated to test this possibility at a muon collider.

In case (ii), one focuses on the theoretically clean lepton flavor universality tests and the $B_s \to \mu^+ \mu^-$ branching ratio. As shown in the lower plot of figure~\ref{fig:more_fits}, this results in a best-fit region that is fully compatible with the SM expectation. A multivariate Gaussian approximation gives
\begin{equation} \label{eq:C9C10_muon_specific}
  \Delta C_9^\mu = -0.28 \pm 0.33 ~,\quad \Delta C_{10}^\mu = -0.07 \pm 0.22~,
\end{equation}
with a larger positive error correlation of $\rho = +86\%$.
The best fit agrees with the SM at $0.8 \sigma$.

For comparison with the sensitivity of a muon collider, we also consider the sensitivity projections of rare $B$ decays after the high-luminosity phase of the LHC. The uncertainties of the current measurements of the $B_s \to \mu^+ \mu^-$ branching ratio are still statistically dominated, and one can expect significant improvements from ATLAS, CMS, and LHCb~\cite{ATLAS:2018cur, LHCb:2021vsc, CMS:2022dbz}. We assume that the measured $B_s \to \mu^+ \mu^-$ branching ratio will coincide with the SM prediction and use an experimental uncertainty of $\pm 0.10 \times 10^{-9}$, a factor 3 reduction in uncertainty compared to the current world average~\cite{HeavyFlavorAveragingGroup:2022wzx}, commensurate with an order of magnitude increase in statistics. Concerning the CKM input for the corresponding theory prediction, one can expect that the CKM matrix element $|V_{cb}|$ will be known with percent level precision from Belle II~\cite{Belle-II:2022cgf}. For our projection, we conservatively assume that the uncertainty on $|V_{cb}|$ will be half the one quoted currently by the PDG~\cite{ParticleDataGroup:2022pth} $|V_{cb}| = (40.8 \pm 1.4)\times 10^{-3} \to |V_{cb}| = (40.8 \pm 0.7)\times 10^{-3}$.

Future LHCb measurements of $R_K$ and $R_{K^*}$ are expected to reach uncertainties at the percent level~\cite{Bifani:2018zmi}. We assume that the measured future central values of $R_K$ and $R_{K^*}$ will coincide with the SM prediction of 1.0 with uncertainties that are a factor of 5 better than those quoted in the most recent analysis~\cite{LHCb:2022qnv, LHCb:2022zom}. This corresponds approximately to the expected uncertainties quoted in~\cite{LHCb:2018roe}.

The constraining power of the $b \to s \mu \mu$ branching ratios and angular observables is already limited by theory uncertainties. To be conservative, we assume no significant improvement in those observables. 

The corresponding projected $1\sigma$ and $2\sigma$ contours are shown in the plots of figure~\ref{fig:more_fits} by the black dashed contours. Gaussian approximations give in case (i)
\begin{equation} \label{eq:C9C10_future_universal}
  \Delta C_9^\text{univ.} = -0.81 \pm 0.20 ~,\quad \Delta C_{10}^\text{univ.} = 0.02 \pm 0.10~,
\end{equation}
with an error correlation of $\rho = -18\%$.  
In case (ii), we find
\begin{equation} \label{eq:C9C10_future_muon_specific}
  \Delta C_9^\mu = 0.00 \pm 0.12 ~,\quad \Delta C_{10}^\mu = 0.00 \pm 0.09~,
\end{equation}
with a large positive error correlation of $\rho = +92\%$.

In section~\ref{sec:SentivityProjections} below, we consider two scenarios: 

On the one hand, we will use the best-fit point from~\eqref{eq:C9C10_benchmark} as a new physics benchmark for lepton universal Wilson coefficients. We will discuss how well a high-energy muon collider can probe such a scenario and compare it to the precision of a future $B$ decay fit~\eqref{eq:C9C10_future_universal}. We note that in contrast to the $b \to s\mu\mu$ observables, bottom-strange production at a muon collider is not significantly affected by long-distance hadronic uncertainties; 

On the other hand, we will assume the absence of new physics and compare the muon collider sensitivity to muon-specific Wilson coefficients to the current and expected sensitivity from rare $B$ decays~\eqref{eq:C9C10_muon_specific} and \eqref{eq:C9C10_future_muon_specific}.

\section{Bottom-Strange Production at a Muon Collider}\label{sec:mumubs}

In addition to a Higgs pole run at a center of mass energy of $\sqrt{s} = 125$\,GeV, a future muon collider is proposed to run at several high energies, including $\sqrt{s} = 3$\,TeV, $\sqrt{s} = 10$\,TeV, and even $\sqrt{s} = 30$\,TeV~\cite{Aime:2022flm, Black:2022cth, Accettura:2023ked}.
Under the assumption that the scale of new physics is sufficiently larger than the center of mass energy, the cross-section for bottom-strange production at a muon collider, $\mu^+ \mu^- \to b \bar s$ or $\bar b s$, can be computed model independently in the effective formalism provided by the Hamiltonian~\eqref{eq:effectiveHamiltonian}. To compute the cross-section, we used~\verb|FeynCalc|~\cite{MERTIG1991345, Shtabovenko:2016sxi, Shtabovenko:2020gxv} and took into account generic polarization fractions of the muon beams. 

We find that the differential cross-sections can be expressed as
\begin{multline} \label{eq:sigma_bsbar}
    \frac{d \sigma(\mu^+\mu^- \to b \bar s)}{dz} = \frac{3}{16} \sigma(\mu^+\mu^- \to b s) \\
    \times \Big( \frac{4}{3} F_S + (1-F_S)\big(1+z^2\big) + \frac{8}{3} z A_\text{FB} \Big) ~,
\end{multline}
\begin{multline} \label{eq:sigma_bbars}
    \frac{d\sigma(\mu^+\mu^- \to \bar b s)}{dz}  = \frac{3}{16} \sigma(\mu^+\mu^- \to b s) \\
    \times \Big( \frac{4}{3} F_S + (1-F_S)\big(1+z^2\big) - \frac{8}{3} z A_\text{FB} \Big) ~,
\end{multline}
where $z = \cos\theta$ with $\theta$ the angle between the $\mu^-$ beam and the $b$ or $\bar b$, respectively.
In the equations above, we have expressed the differential cross-sections in terms of the total cross-section
\begin{multline}
\sigma(\mu^+\mu^- \to b s) = 2 \sigma(\mu^+\mu^- \to b \bar s) = 2 \sigma(\mu^+\mu^- \to \bar b s) \\
= \frac{G_F^2 \alpha^2}{8 \pi^3} |V_{tb} V_{ts}^*|^2 s \Big( \frac{3}{4} a_0 + a_2 \Big) ~,
\end{multline}
as well as the forward-backward asymmetry of the bottom quark, $A_\text{FB}$, and the fraction of the cross-section that originates from scalar or pseudoscalar operators, $F_S$,
\begin{equation} \label{eq:AFB}
 A_\text{FB} = \frac{3 a_1}{3 a_0 + 4 a_2} ~, \quad F_S = \frac{3 a_0}{3 a_0 + 4 a_2} ~.
\end{equation}
The coefficients $a_0$, $a_1$, and $a_2$ are given by the following combinations of Wilson coefficients
\begin{multline}
 a_0 = (1 + P_+ P_-) \Big( |C_S|^2 + |C_P|^2 + |C'_S|^2 + |C'_P|^2 \Big) \\
 + 2 ( P_- + P_+) \Big( \mathfrak{Re}(C_S C_P^*) + \mathfrak{Re}(C'_S C_P^{\prime *})  \Big)  ~,
\end{multline}
\begin{multline}
 a_1 = ( P_+ - P_-) \Big( |\Delta C_9|^2 + |\Delta C_{10}|^2 - |C'_9|^2 - |C'_{10}|^2 \Big) \\
 - 2 (1 - P_+ P_-) \Big( \mathfrak{Re}(\Delta C_9 \Delta C_{10}^*) - \mathfrak{Re}(C'_9 C_{10}^{\prime *})  \Big)  ~,
\end{multline}
\begin{multline}
 a_2 = (1 - P_+ P_-) \Big( |\Delta C_9|^2 + |\Delta C_{10}|^2 + |C'_9|^2 + |C'_{10}|^2 \Big) \\ 
 + 2 ( P_- - P_+) \Big( \mathfrak{Re}(\Delta C_9 \Delta C_{10}^*) + \mathfrak{Re}(C'_9 C_{10}^{\prime *})  \Big)  ~.
\end{multline}
where for better readability, we dropped the lepton superscript ``$\mu$'' on the Wilson coefficients, c.f. eq.~\eqref{eq:effectiveHamiltonian}.
The beam polarizations $P_\pm \in [-1,1]$ specify the fraction of polarized $\mu^+$ and $\mu^-$, respectively, with $P_\pm = +1 (-1)$ indicating purely right-handed (left-handed) beams. The unpolarized limit is restored by setting $P_\pm = 0$. 

In the absence of beam polarization, the total cross-section simplifies to
\begin{multline}
\sigma(\mu^+\mu^- \to b s) =  \\
= \frac{G_F^2 \alpha^2}{8 \pi^3} |V_{tb} V_{ts}^*|^2 s \Big[ |\Delta C_9|^2 + |\Delta C_{10}|^2 + |C'_9|^2 + |C'_{10}|^2 \\
+ \frac{3}{4} \Big( |C_S|^2 + |C_P|^2 + |C'_S|^2 + |C'_P|^2 \Big) \Big] ~.
\end{multline}

As expected from dimensional analysis, the signal cross section grows linearly with the center of mass energy squared, $s$. Standard Model background processes (see the discussion in section~\ref{sec:collider}), are expected to fall with the center of mass energy. At sufficiently high center of mass energies, a muon collider will thus be able to detect a new physics signal in the benchmark scenario~\eqref{eq:C9C10_benchmark}.  
If a new physics signal is established, measurements of the forward-backward asymmetry, $A_\text{FB}$ provide further information about the relative size of $C_9^{(\prime)}$ and $C_{10}^{(\prime)}$ Wilson coefficients. Note that the forward-backward asymmetry enters the differential $\mu^+\mu^- \to b \bar s$ and  $\mu^+\mu^- \to b \bar s$ cross sections with opposite sign, c.f. equations~\eqref{eq:sigma_bsbar} and~\eqref{eq:sigma_bbars}. A measurement of $A_\text{FB}$ thus requires charge tagging of the $b$ jet.

Both the cross-section and the forward-backward asymmetry are affected by the degree of muon beam polarization. 
The muons are produced from pion decay, and the outgoing muon is fully polarized in the center-of-mass frame of a decaying pion. In the lab frame, on the other hand, the polarization depends on the
decay angle and pion energy and is typically around 20\%~\cite{Ankenbrandt:1999cta}.
Higher polarization can be achieved if muons from forward pion decays are selected. This comes at the expense of luminosity. For example, a polarization of $\sim 50\%$ might be achieved for a decrease in luminosity by a factor of $\sim 4$~\cite{Ankenbrandt:1999cta}. 

As can be seen from the equations for the cross-section and the forward-backward asymmetry above, the beam polarization does have an impact. In the numerical analysis discussed below, we will consider as representative cases unpolarized muon beams, as well as muon beams with $+ 50\%$ polarization.

\section{Background Processes}\label{sec:collider}

Various background processes contribute to a $\mu^+ \mu^- \to b s$ signal at a muon collider.
On the one hand, there is an irreducible SM background that is suppressed by the Glashow–Iliopoulos–Maiani (GIM) mechanism. On the other hand, there are reducible backgrounds from di-jet production $\mu^+ \mu^- \to jj$ where one of the jets is incorrectly flavor tagged, as well as backgrounds from processes with missing energy, $\mu^+ \mu^- \to b s +\text{ E \!\!\!\!\!/}$.
We detail the various types of backgrounds in the following. Example Feynman diagrams are shown in figures~\ref{fig:diagrams_1},~\ref{fig:diagrams_2}, and~\ref{fig:diagrams_3}.

\subsection{SM loop contribution}

\begin{figure}[tb]
 \centering
  \includegraphics[width=0.22\textwidth]{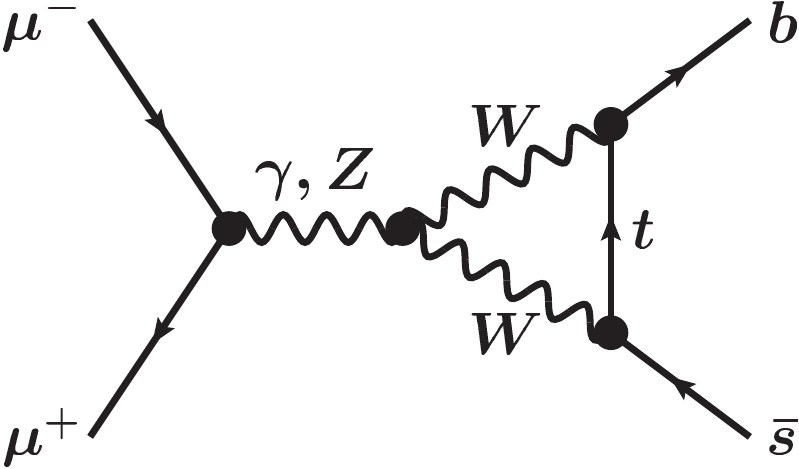} \qquad
  \includegraphics[width=0.20\textwidth]{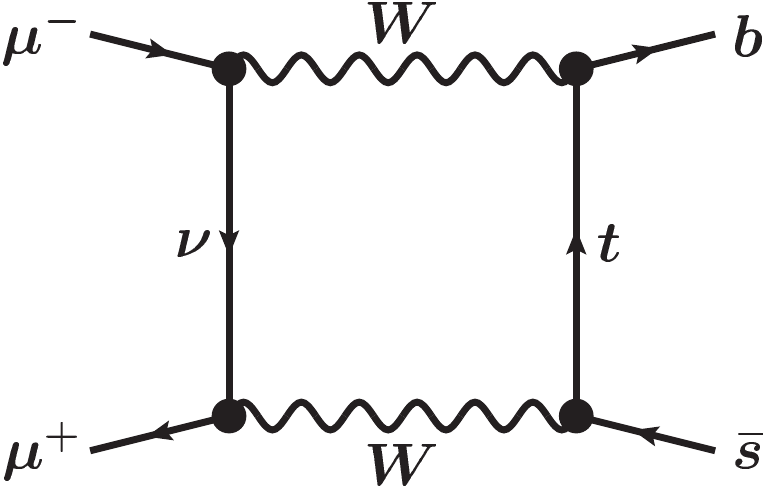} 
  \caption{Example Feynman diagrams for the irreducible one-loop SM background $\mu^+ \mu^- \to b s$.}
  \label{fig:diagrams_1}
\end{figure}

The irreducible SM contribution to the  $\mu^+\mu^- \to b s$ cross-section arises at the one-loop level and is, as mentioned above, GIM suppressed. Example diagrams are shown in figure~\ref{fig:diagrams_1}. At a high-energy muon collider, this SM loop contribution cannot be described by a contact interaction but requires a calculation with dynamical top quarks, $W$ bosons, and $Z$ bosons. We have calculated this cross section for an arbitrary center of mass energy $\sqrt{s}$ using \verb|FeynArts|~\cite{Hahn:2000kx} and \verb|FormCalc|~\cite{Hahn:1998yk}. For a large $\sqrt{s} \gg m_t, m_W, m_Z$, we find that the cross-section falls with $1/s$, or more precisely
\begin{equation}
\sigma_\text{bg}^\text{loop} \propto \frac{G_F^2 m_t^4 \alpha^2}{128 \pi^3} |V_{tb} V_{ts}^*|^2 \frac{1}{s} ~.
\end{equation}
This turns out to be completely negligible at a multi-TeV muon collider.

\subsection{Mistagged di-jet events}

\begin{figure}[tb]
 \centering
  \includegraphics[width=0.19\textwidth]{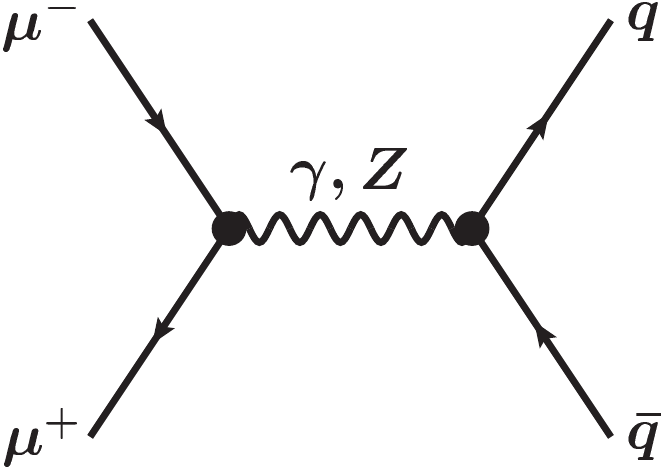}
  \caption{Feynman diagram for di-jet production, $\mu^+ \mu^- \to q \bar q$.}
  \label{fig:diagrams_2}
\end{figure}

A much more important source of background stems from mistagged di-jet events from the diagram shown in figure~\ref{fig:diagrams_2}. We consider the production of $b$ jets, $\mu^+ \mu^- \to b \bar b$, in which one $b$-jet is misidentified as a light jet, as well as $\mu^+ \mu^- \to c \bar c$ and $\mu^+ \mu^- \to q \bar q$ events with light quarks $q = u,d,s$, where one of the charm or light quark jets is identified as a $b$-jet. We analytically calculated the corresponding di-jet cross sections at tree level. We assume that top tagging at a muon collider is sufficiently accurate such that $\mu^+ \mu^- \to t \bar t$ events do not give a relevant background. The corresponding background cross section from di-jets that we consider is therefore
\begin{equation}
  \sigma_\text{bg}^\text{jj} = \sum_{q = u,d,s,c,b} 2 \epsilon_q (1-\epsilon_q)~ \sigma(\mu^+ \mu^- \to q \bar q) ~,
\end{equation}
where $\epsilon_b$ is the $b$-tag efficiency and $\epsilon_{u,d,s,c}$ the probabilities that a charm or light quark jet is misidentified as a $b$-jet. 
For the numerical analysis we follow~\cite{Huang:2021biu} and adopt the values: $\epsilon_b = 70\%$, $\epsilon_c = 10\%$, and $\epsilon_u = \epsilon_d = \epsilon_s = 1\%$. These values are comparable to those that are currently achieved by the ATLAS and CMS experiments at the LHC for jets with transverse momentum up to a few hundred GeV~\cite{CMS:2017wtu, ATLAS:2017bcq}. The performance of traditional flavor taggers decreases significantly for a jet $p_T$ in the multi-TeV regime. However, novel tagging techniques~\cite{PerezCodina:2631478} should improve the performance for multi-TeV jets to the level quoted above.

\subsection{Di-jet events from vector boson fusion}

\begin{figure}[tb]
 \centering
  \includegraphics[width=0.21\textwidth]{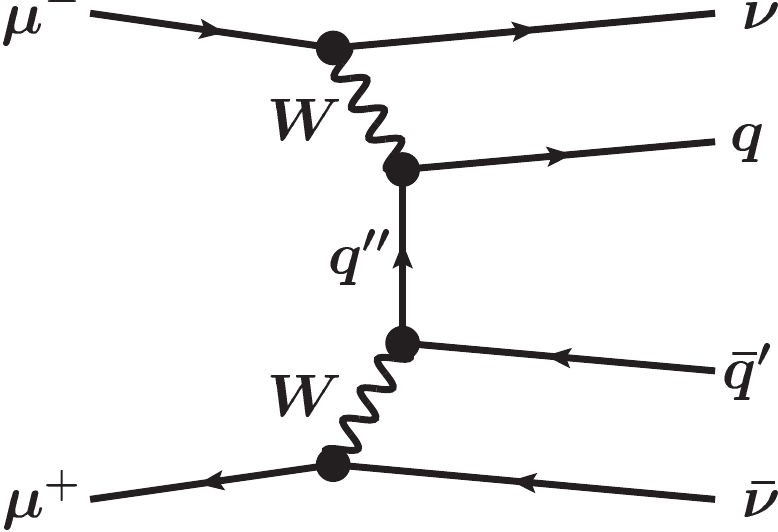} \qquad
  \includegraphics[width=0.21\textwidth]{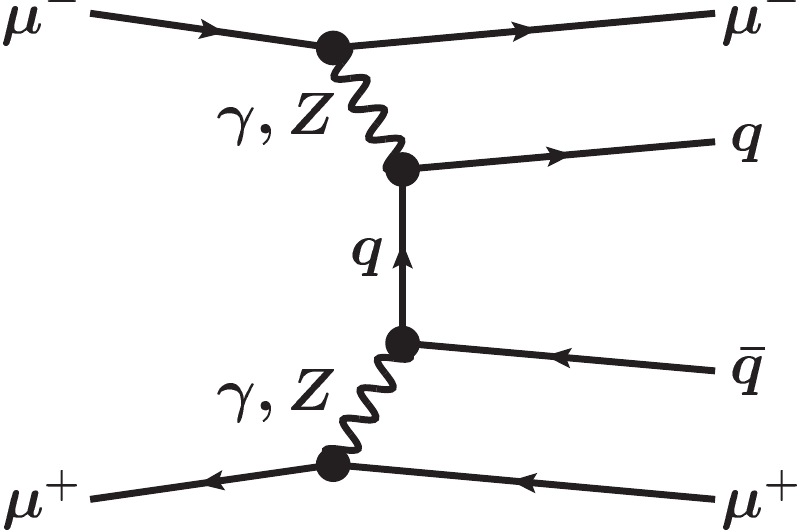} \\[24pt]
  \includegraphics[width=0.21\textwidth]{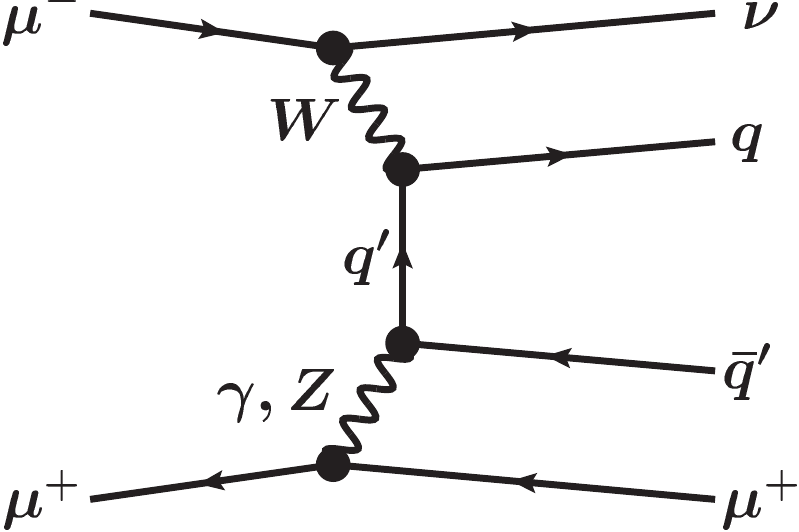} 
  \caption{Example Feynman diagrams for di-jet production from vector boson fusion $\mu^+\mu^- \to q \bar q^\prime \nu \bar\nu$, $\mu^+\mu^- \to q \bar q \mu^+ \mu^-$, and $\mu^+\mu^- \to q \bar q^\prime \mu \nu$.}
  \label{fig:diagrams_3}
\end{figure}

Finally, additional backgrounds come from di-jet production through vector boson fusion in association with forward muons or neutrinos that remain undetected.
The relevant processes are $\mu^+\mu^- \to b \bar b \nu \bar \nu$, $\mu^+\mu^- \to c \bar c \nu \bar \nu$, or $\mu^+\mu^- \to q \bar q \nu \bar \nu$ with mistagged quarks, $\mu^+\mu^- \to b \bar s \nu \bar \nu$, as well as $\mu^+\mu^- \to q \bar q^\prime \mu^+ \mu^-$ and $\mu^+\mu^- \to q \bar q^\prime \mu \nu$ with appropriate quark flavors. Example diagrams can be found in figure~\ref{fig:diagrams_3}.
These processes are potentially relevant as the vector boson fusion cross section grows with the center of mass energy~\cite{Costantini:2020stv}. 
This form of background can be largely removed by cuts on the di-jet invariant mass. For $\mu^+\mu^- \to b s$ signal events, one expects $m_{jj} \simeq \sqrt{s}$, while for vector boson fusion events, one expects a significantly reduced di-jet invariant mass, $m_{jj} < \sqrt{s}$, due to the forward muons or neutrinos carrying away energy. 
A di-jet invariant mass resolution of $\sim 2\%$ for 5~TeV di-jets has been achieved at ATLAS~\cite{ATLAS:2017eqx}. We assume that detectors at a future muon collider will perform at least as well. 
We determine the cross sections of the vector boson fusion background, $\sigma_\text{bg}^\text{VBF}$, using \verb|MadGraph5|~\cite{Alwall:2014hca} to simulate the 4 body final states $\mu^+\mu^- \to q \bar q^\prime \nu \bar \nu$, $\mu^+\mu^- \to q \bar q^\prime \mu \nu$, and $\mu^+\mu^- \to q \bar q^\prime \mu^+ \mu^-$ with all relevant quark flavors. We employ a cut on the di-jet invariant mass of $m_{jj}/\sqrt{s} = 1 \pm 0.04$. Such a cut retains $\simeq 95\%$ of the signal but reduces this background by $5-6$ orders of magnitude, to a subdominant level. 

A more precise calculation of this background could be done by making use of lepton PDFs~\cite{Han:2020uid, Han:2021kes, Ruiz:2021tdt, Garosi:2023bvq}. This is left for future work.

\subsection{Comparison of signal and background}

\begin{figure*}[tb]
 \centering
  \includegraphics[width=0.7\linewidth]{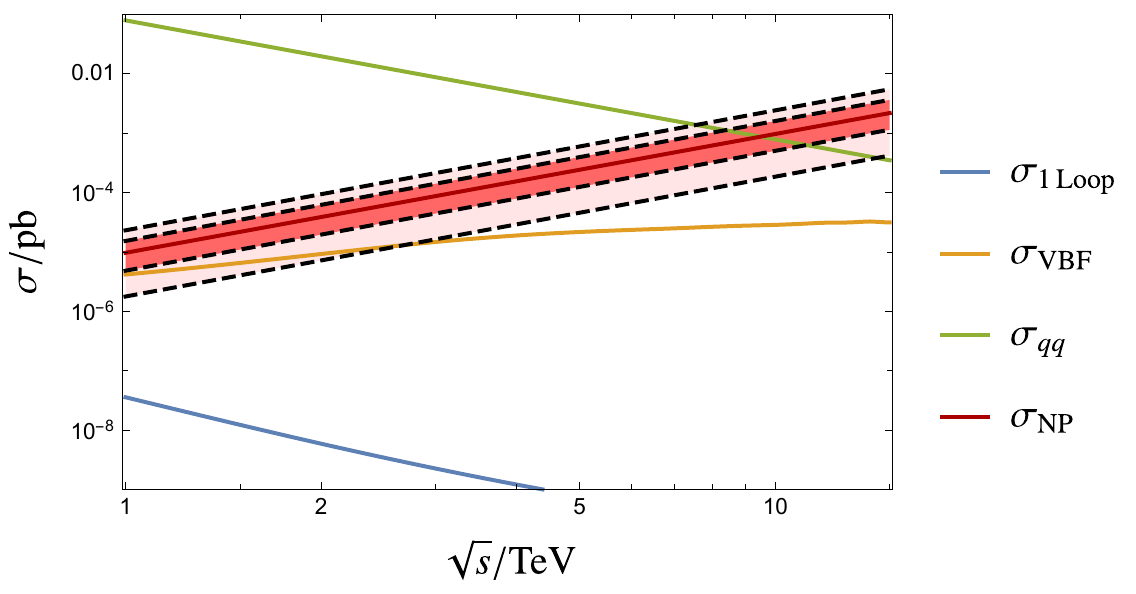} 
  \caption{The cross sections of the $\mu^+\mu^- \to b s$ signal and background processes at a muon collider with a center of mass energy $\sqrt{s}$. The shown cross sections take into account flavor tagging efficiencies and mistag rates as discussed in the text. The signal cross section corresponds to the benchmark point~\eqref{eq:C9C10_benchmark} with $1\sigma$ and $2\sigma$ uncertainties. The muon beams are assumed to be unpolarized.}
  \label{fig:cross_sections}
\end{figure*}

In figure~\ref{fig:cross_sections}, we show as a function of the center of mass energy $\sqrt{s}$ the cross sections of the new physics $\mu^+\mu^- \to b s$ signal, $\sigma_\text{NP}$ (red), and the background processes mentioned above, namely the irreducible SM one-loop contribution, $\sigma_\text{bg}^\text{loop}$ (blue), mistagged di-jets, $\sigma_\text{bg}^{jj}$ (green), and di-jets from vector boson fusion, $\sigma_\text{bg}^\text{VBF}$ (orange).
The shown VBF cross-section only includes the $\mu^+\mu^- \to q \bar q^\prime \nu \bar \nu$ processes. The $\mu^+\mu^- \to q \bar q^\prime \mu \nu$ and $\mu^+\mu^- \to q \bar q^\prime \mu^+ \mu^-$ cross sections are somewhat larger, but we expect that they can be significantly reduced by vetoing muons in the event.
For the signal cross-section, we assume a new physics benchmark as in~\eqref{eq:C9C10_benchmark}, with the red shaded regions indicating the $1\sigma$ and $2\sigma$ uncertainties.  All shown cross-sections take into account the flavor tagging efficiencies and mistag rates.
The muon beams are assumed to be unpolarized. Qualitatively similar results are obtained for polarized muon beams.

At a low center of mass energy, the mistagged di-jet background dominates the signal by orders of magnitude. Among the di-jet backgrounds, the $b\bar b$ final state contributes the most, followed by $c \bar c$ and light quarks. The SM loop background and the background from vector boson fusion (with $m_{jj}$ cut) are subdominant and we will neglect them in the following. As anticipated, the signal cross-section increases with $s$, while the most important background cross-section falls approximately like $1/s$. Signal and background become comparable for a center of mass energy of around 10~TeV. This suggests that a 10~TeV muon collider should be able to observe a non-standard $\mu^+ \mu^- \to bs$ production with high significance.

\section{Sensitivity Projections} \label{sec:SentivityProjections}

Based on the discussion in the previous section, we investigate the sensitivity of a multi-TeV muon collider to the contact interactions in~\eqref{eq:effectiveHamiltonian}. Proposed runs of a muon collider include a center of mass energy of $\sqrt{s} = 6$\,\, TeV with an integrated luminosity of $\mathcal L = 4$\,ab$^{-1}$ and $\sqrt{s} = 10$\,TeV with $\mathcal L = 10$\,ab$^{-1}$~\cite{Black:2022cth}. 

\subsection{Lepton flavor universal new physics benchmark}
\label{sec:universal}

First, we discuss the sensitivity to the new physics benchmark point~\eqref{eq:C9C10_benchmark}, which corresponds to lepton flavor universal new physics. As discussed in section~\ref{sec:GlobalFit}, this new physics benchmark is unconstrained by the lepton flavor universality ratios $R_K$ and $R_{K^*}$ and motivated by the anomalously low branching ratios of $b \to s \mu\mu$ decays and the angular distribution of $B \to K^* \mu\mu$. We stress again that these hints for new physics rely on the modeling of hadronic effects in rare $b$ decays and might be due to underestimated theory uncertainties. A completely independent cross-check at a muon collider would therefore be more than welcome.

\setlength{\tabcolsep}{6pt}
\renewcommand{\arraystretch}{2.0}
\begin{table}[tbh]
\centering
\begin{tabular}{cccc}
\hline\hline
 & 6\,TeV, 4\,ab$^{-1}$ & 10\,TeV, 1\,ab$^{-1}$ & 10\,TeV, 10\,ab$^{-1}$ \\
\hline\hline
$N_\text{tot}$ & $10050 \pm 220$ & $1,740 \pm 50$ & $17,400 \pm 220$ \\
$N_\text{bg}$ & $8670 \pm 220$ & $780 \pm 42$ & $7,800 \pm 200$ \\
$N_\text{sig}$ & $1380 \pm 310$ & $960 \pm 68$ & $9,600 \pm 300$ \\
\hline\hline
\end{tabular}
\caption{Expected $\mu^+\mu^- \to b s$ event numbers at different configurations of a muon collider with unpolarized muon beams.}
\label{tab:event_numbers_unpolarized}
\end{table}

With the chosen benchmark point, we obtain the expected $\mu^+\mu^- \to b s$ event numbers for unpolarized muon beams summarized in table~\ref{tab:event_numbers_unpolarized}.
The numbers include the flavor tagging efficiencies and mistag rates discussed above. Explicitly, this means
\begin{eqnarray}
N_\text{bg} &=& \mathcal L \times \sum_{q = u,d,s,c,b} 2 \epsilon_q (1-\epsilon_q)~ \sigma(\mu^+ \mu^- \to q \bar q) ~, \\
N_\text{tot} &=& N_\text{bg} + \mathcal L \times \epsilon_b (1-\epsilon_s)~ \sigma(\mu^+ \mu^- \to b s) ~.
\end{eqnarray}
The quoted uncertainties on the total event numbers, $N_\text{tot}$, and the background event numbers, $N_\text{bg}$, include the statistical as well as a 2\% systematic uncertainty added in quadrature. In all cases, the total number of events is significantly above the background prediction.
The number of signal events is determined from $N_\text{sig} = N_\text{tot} - N_\text{bg}$, with the errors added in quadrature.
Based on these numbers, we expect that the signal cross section can be measured with a precision of $\sim 22\%$ at 6~TeV with 4\,ab$^{-1}$, $\sim 7\%$ at 10~TeV with 1\,ab$^{-1}$, and $\sim 3\%$ at 10~TeV with 10\,ab$^{-1}$.\footnote{Note that the expected precision is significantly better compared to the preliminary results we reported in the whitepaper~\cite{Altmannshofer:2022xri}. This is due to the change of the signal benchmark point \eqref{eq:C9C10_benchmark} motivated by the new $R_K$, $R_{K^*}$ results~\cite{LHCb:2022qnv,LHCb:2022zom}.}

For a beam polarization of $P_- = - P_+ = 50\%$, we analogously find the event numbers in table~\ref{tab:event_numbers_polarized}. Both background and signal event numbers are slightly smaller in this case. This results in a comparable expected precision on the signal cross-section.

\begin{table}[tbh]
\centering
\begin{tabular}{cccc}
\hline\hline
 & 6\,TeV, 4\,ab$^{-1}$ & 10\,TeV, 1\,ab$^{-1}$ & 10\,TeV, 10\,ab$^{-1}$ \\
\hline\hline
$N_\text{tot}$ & $7890 \pm 180$ & $1,490 \pm 50$ & $14,580 \pm 190$ \\
$N_\text{bg}$ & $6610 \pm 180$ & $600 \pm 40$ & $5,947 \pm 160$ \\
$N_\text{sig}$ & $1280 \pm 250$ & $890 \pm 60$ & $8905 \pm 250$ \\
\hline\hline
\end{tabular}
\caption{Expected $\mu^+\mu^- \to b s$ event numbers at different configurations of a muon collider with a beam polarization of $P_- = - P_+ = 50\%$.}
\label{tab:event_numbers_polarized}
\end{table}

\begin{figure*}[tbh]
 \centering
  \includegraphics[width=0.46\textwidth]{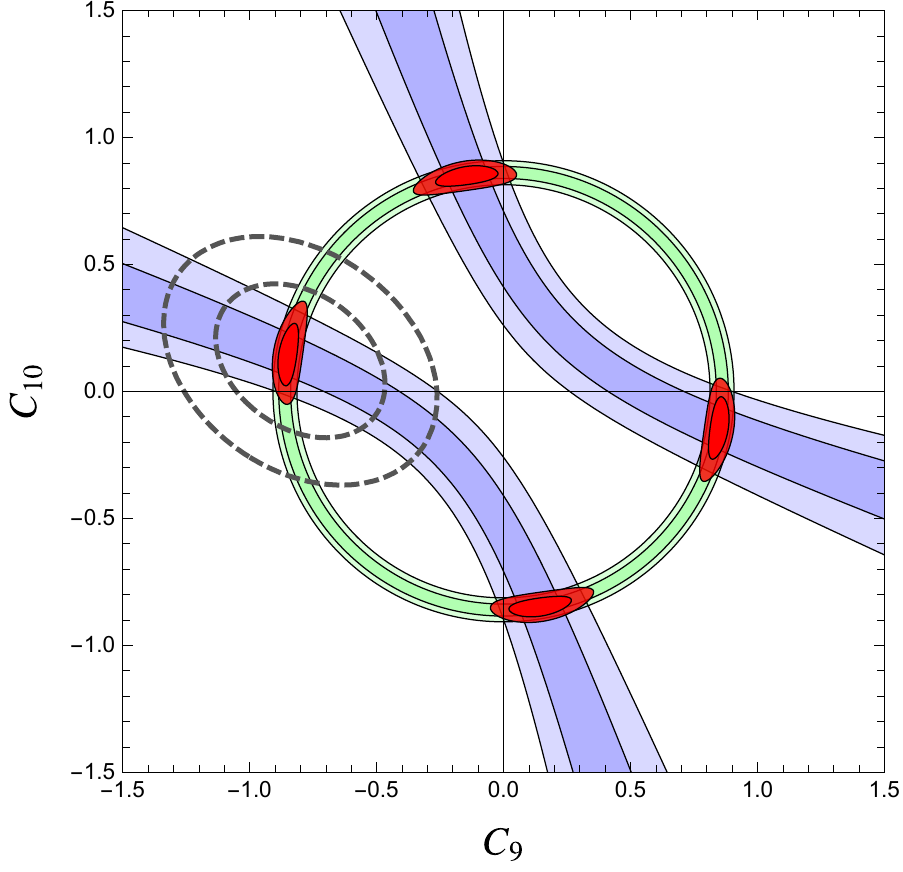} \enspace \enspace
  \includegraphics[width=0.46\textwidth]{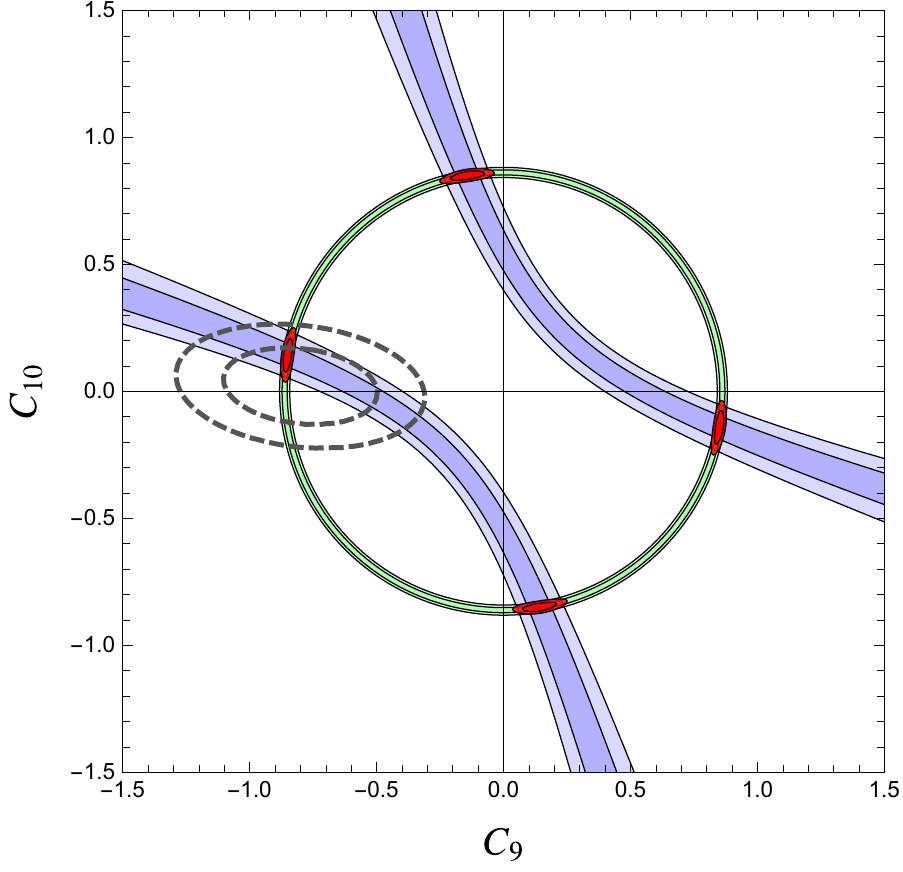}
  \caption{Sensitivity of a 10~TeV muon collider with unpolarized beams in the $\Delta C_9^\text{univ.}$ vs. $\Delta C_{10}^\text{univ.}$ plane, assuming the new physics benchmark point in \eqref{eq:C9C10_benchmark}.
  Shown in green (blue) is the region that can be determined by a measurement of the $\mu^+\mu^- \to b s$ cross section (the forward-backward asymmetry). The combination is in red. The left (right) plot assumes an integrated luminosity of 1~ab$^{-1}$ (10~ab$^{-1}$). The dashed black lines are the current best-fit region from rare $B$ decays \eqref{eq:C9C10_benchmark} (left plot) or the expected region after the HL-LHC and Belle II \eqref{eq:C9C10_future_universal} (right plot).}
  \label{fig:C9_vs_C10_benchmark}
\end{figure*}
\begin{figure*}[tbh]
 \centering
  \includegraphics[width=0.46\textwidth]{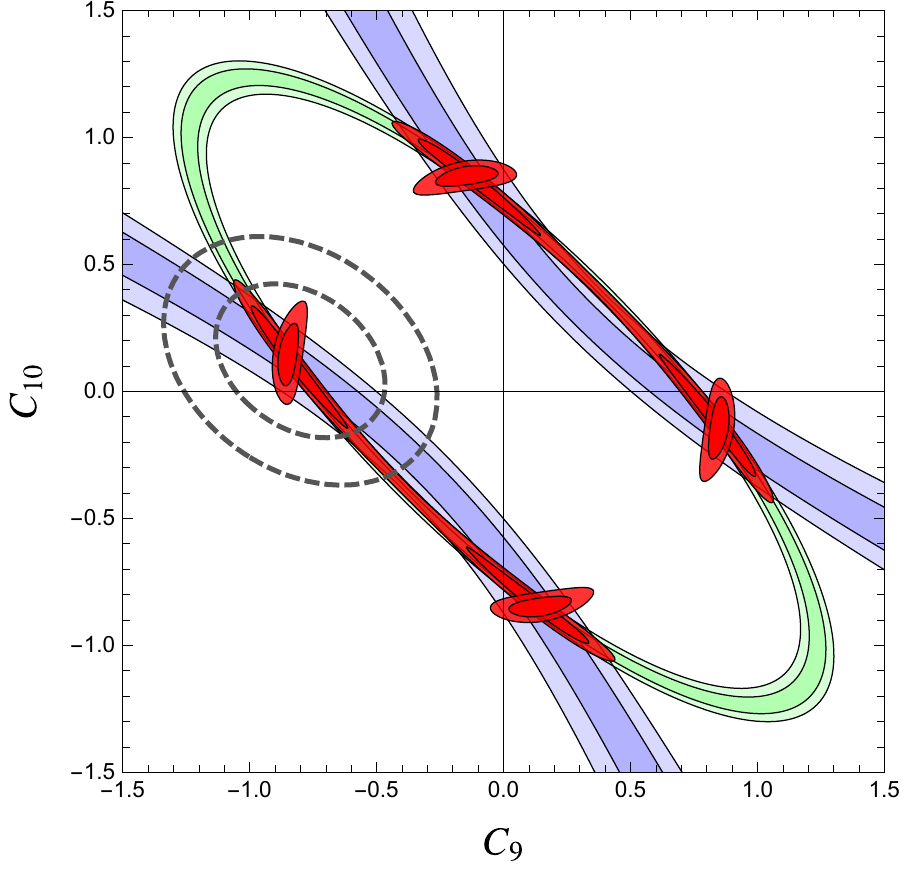} \enspace \enspace
  \includegraphics[width=0.46\textwidth]{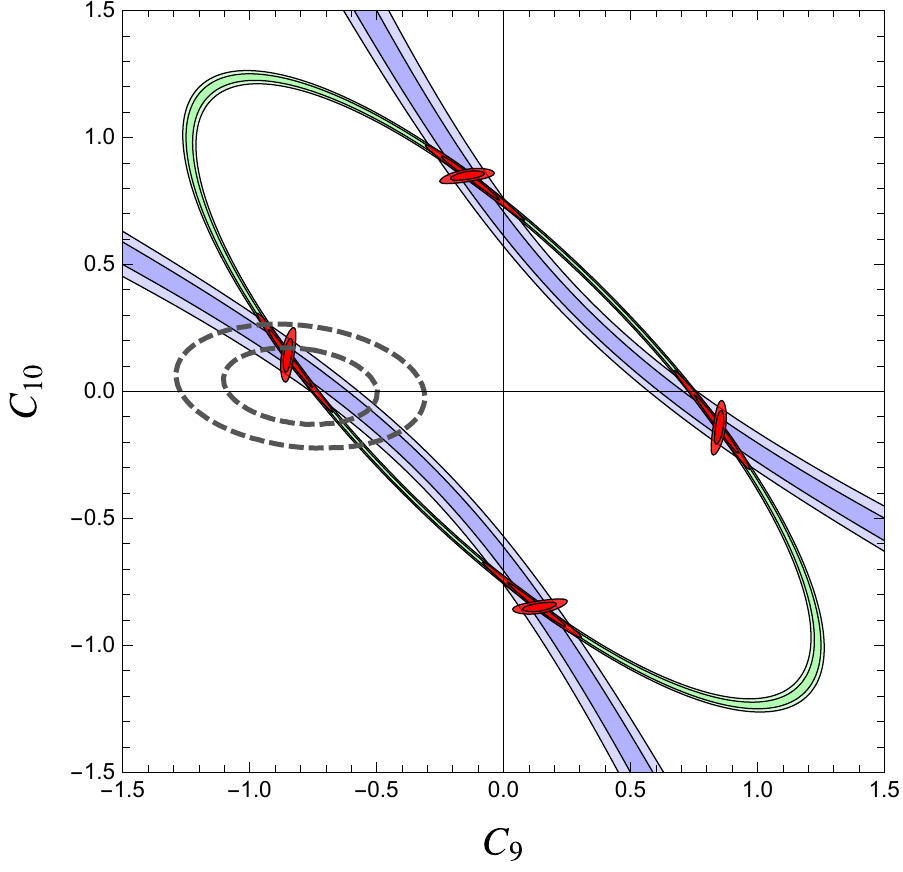}
  \caption{Same as the plots in figure~\ref{fig:C9_vs_C10_benchmark}, but with a muon beam polarization of $P_- = - P_+ = 50\%$. The combination of the unpolarized best-fit regions are overlaid on the polarized regions, highlighting the complementarity of the polarized and unpolarized beams.}
  \label{fig:C9_vs_C10_benchmark_polarized}
\end{figure*}

The $1\sigma$ and $2\sigma$ constraints in the $C_9 - C_{10}$ plane from the expected measurement of the $\mu^+ \mu^- \to bs$ cross-section are shown in green in figures~\ref{fig:C9_vs_C10_benchmark} and~\ref{fig:C9_vs_C10_benchmark_polarized}.

The forward-backward asymmetry introduced in \eqref{eq:sigma_bsbar} and \eqref{eq:sigma_bbars} provides complementary information about the new physics. As mentioned in section~\ref{sec:mumubs}, a measurement of the forward-backward asymmetry requires charge tagging in addition to flavor tagging. We estimate the expected precision of a $A_\text{FB}$ measurement by splitting the events into forward and backward categories.
A better precision could likely be obtained by performing an unbinned maximum likelihood fit to the angular distributions in \eqref{eq:sigma_bsbar} and \eqref{eq:sigma_bbars}. This is, however, beyond the scope of this work. Denoting the charge tagging efficiency by $\epsilon_\pm$, and implicitly including flavor tagging efficiencies, the expected number of observed forward and backward signal events is given by
\begin{eqnarray} \label{eq:N_F_observed}
        N^{\text{F},b \overline{s}}_{\text{sig,}\, \text{obs}} &=& \epsilon_\pm \, N^{\text{F},b \overline{s}}_\text{sig} + (1-\epsilon_\pm)\, N^{\text{F}, s \overline{b}}_\text{sig} ~, \\   \label{eq:N_F_observed}      N^{\text{B},b \overline{s}}_{\text{sig,}\, \text{obs}} &=& \epsilon_\pm \, N^{\text{B},b \overline{s}}_\text{sig} + (1-\epsilon_\pm)\, N^{\text{B}, s \overline{b}}_\text{sig} ~, \\
        N^{\text{F},s \overline{b}}_{\text{sig,}\, \text{obs}} &=& \epsilon_\pm \, N^{\text{F},s \overline{b}}_\text{sig} + (1-\epsilon_\pm)\, N^{\text{F}, b \overline{s}}_\text{sig} ~, \\   \label{eq:N_F_observed}      N^{\text{B},s \overline{b}}_{\text{sig,}\, \text{obs}} &=& \epsilon_\pm \, N^{\text{B},s \overline{b}}_\text{sig} + (1-\epsilon_\pm)\, N^{\text{B}, b \overline{s}}_\text{sig} ~,
\end{eqnarray}
and analogously for background events from flavor mistags. The observed total forward-backward asymmetry, $A_\text{FB}^\text{obs}$, is then given by
\begin{equation} \label{eq:AFB_observed}
    A_\text{FB}^\text{obs} = \frac{N^\text{F}_\text{obs} - N^\text{B}_\text{obs}}{N^\text{F}_\text{obs} + N^\text{B}_\text{obs}} ~,
\end{equation}
where the observed event numbers are the sum of signal and background and take into account charge tagging and flavor tagging
\begin{eqnarray}
 N_\text{obs}^\text{F} &=& N^{\text{F},b \overline{s}}_{\text{sig,}\, \text{obs}} + N^{\text{B},\overline{b} s}_{\text{sig,}\, \text{obs}} + N^{\text{F},b \overline{s}}_{\text{bg,}\, \text{obs}} + N^{\text{B},\overline{b} s}_{\text{bg,}\, \text{obs}} ~, \\  
 N_\text{obs}^\text{B} &=& N^{\text{B},b \overline{s}}_{\text{sig,}\, \text{obs}} + N^{\text{F},\overline{b} s}_{\text{sig,}\, \text{obs}} + N^{\text{B},b \overline{s}}_{\text{bg,}\, \text{obs}} + N^{\text{F},\overline{b} s}_{\text{bg,}\, \text{obs}} ~.
\end{eqnarray}
Alternatively, $A_\text{FB}^\text{obs}$ can be expressed as the following combination of the truth level signal forward-backward asymmetry, $A_\text{FB}$ as given in \eqref{eq:AFB}, and the truth level background forward-backward asymmetry, $A_\text{FB}^\text{bg}$,
\begin{equation} \label{eq:AFB_observed_2}
    A_\text{FB}^\text{obs} 
    = (2\epsilon_\pm - 1) \Big( \frac{N_\text{sig}}{N_\text{tot}} A_\text{FB} + \frac{N_\text{bg}}{N_\text{tot}} A_\text{FB}^\text{bg} \Big) ~,
\end{equation}
Assuming unpolarized muon beams, the signal forward-backward asymmetry of the chosen benchmark point \eqref{eq:C9C10_benchmark} is $A_\text{FB} \simeq 0.24$, while for the background we find $A_\text{FB}^\text{bg} \simeq 0.62$. For polarized beams these values change to $A_\text{FB} \simeq -0.48$ and $A_\text{FB}^\text{bg} \simeq 0.59$.

We assume that the $b$-jet charge tagging performance of a future muon collider will be comparable to that achieved at LEP, $\epsilon_\pm = 70\%$~\cite{DELPHI:2004wzo}. The imperfect charge tagging washes out the observed forward-backward asymmetry by a factor of $(2\epsilon_\pm - 1) = 0.4$, as shown in \eqref{eq:AFB_observed_2}.

The uncertainty on the observed forward-backward asymmetry can be estimated from~\eqref{eq:AFB_observed}. Treating the number of forward and backward events as independent, we find
\begin{equation}
\delta A_\text{FB}^\text{obs} = \frac{2}{N_\text{tot}^2}\sqrt{(N^\text{F}_\text{obs})^2(\delta N^\text{B}_\text{obs})^2+ (N^\text{B}_\text{obs})^2(\delta N^\text{F}_\text{obs})^2} ~.
\end{equation}
We expect that this slightly overestimates the uncertainty.
In the determination of the $\delta N^\text{F}_\text{obs}$ and $\delta N^\text{B}_\text{obs}$, we take into account the statistical as well as a 2\% systematic uncertainty.

We find expected measurements of the total forward-backward asymmetry $A_\text{FB}^\text{obs} = (22.7 \pm 1.7)\%$ at 6 TeV with 4~ab$^{-1}$, $A_\text{FB}^\text{obs} = (16.4 \pm 2.9)\%$ at 10 TeV with 1~ab$^{-1}$, and $A_\text{FB}^\text{obs} = (16.4 \pm 1.6)\%$ at 10 TeV with 10~ab$^{-1}$.

The forward-backward asymmetry is highly complementary to the cross-section and leads to orthogonal constraints in the $C_9 - C_{10}$ plane, presented in the blue regions in figures~\ref{fig:C9_vs_C10_benchmark} and~\ref{fig:C9_vs_C10_benchmark_polarized}.

The combination of cross-section and forward-backward asymmetry is shown in red.
For comparison, the dashed black contours in the plots on the left-hand side show the $1\sigma$ and $2\sigma$ best-fit region of the current rare $B$ decay fit from~\eqref{eq:C9C10_benchmark}. 
The dashed black contours in the plots on the right-hand side correspond to the $1\sigma$ and $2\sigma$ region of the projection~\eqref{eq:C9C10_future_universal}.

The plots illustrate that a 10~TeV muon collider could establish a new physics signal with remarkable precision. Interestingly, a muon collider would select regions in the new physics parameter space with a four-fold degeneracy. Combining the information from the muon collider with the information from rare $B$ decays allows one to uniquely identify the new physics. The best-fit new physics region determined by a 10 TeV muon collider with unpolarized beams and 1\,ab$^{-1}$ of data that is compatible with the rare $B$ decay data is 
\begin{equation} \label{eq:C9C10_1ab}
  \Delta C_9^\text{univ.} = -0.81 \pm 0.03 ~,\quad \Delta C_{10}^\text{univ.} = 0.12 \pm 0.08~,
\end{equation}
with an error correlation of $\rho = +40\%$.
For 10\,ab$^{-1}$, this further improves to 
\begin{equation} \label{eq:C9C10_10ab}
  \Delta C_9^\text{univ.} = -0.81 \pm 0.01 ~,\quad \Delta C_{10}^\text{univ.} = 0.12 \pm 0.04~.
\end{equation}
with an error correlation of $\rho = +53\%$. Going to even higher luminosity has little impact as the precision starts to be limited by systematic uncertainties. 
We note that the expected muon collider results are much more precise than the expected precision from rare $B$ decays alone~\eqref{eq:C9C10_future_universal}. 

The sensitivity of a muon collider could be improved even further if multiple runs with different beam polarizations were an option. As shown in the plots of figure~\ref{fig:C9_vs_C10_benchmark_polarized}, beam polarization does shape the best-fit regions that are selected in the new physics parameter space. As the various operators in~\eqref{eq:effectiveHamiltonian} correspond to different linear combinations of muon chiralities, changing the polarizations of the muon beams also changes the sensitivity to different types of operators and would allow one to narrow down the parameter space further.

\subsection{Constraints on muon-specific new physics} \label{sec:muon-specific}

In the absence of new physics, a high-energy muon collider can constrain the size of the Wilson coefficients in Eq.~\eqref{eq:effectiveHamiltonian}. Switching on one Wilson coefficient at a time and demanding that the number of $\mu^+ \mu^- \to b s$ signal events does not exceed the $2\sigma$ uncertainty of the background, we find for unpolarized muon beams 
\begin{equation} \label{eq:vector_bound}
    |C_\text{vector}| < \begin{cases} 0.46 ~~~ @ ~6~\text{TeV}~,~~~ 4~\text{ab}^{-1} \\
    0.22 ~~~ @ ~10~\text{TeV} ~,~~ 1~\text{ab}^{-1} \\ 0.17 ~~~ @ ~10~\text{TeV} ~,~~ 10~\text{ab}^{-1} \end{cases} ~,
\end{equation}
for the vector Wilson coefficients $C_\text{vector} = \Delta C_9^\mu$, $\Delta C_{10}^\mu$, $C_9^{\prime \mu}$, or $C_{10}^{\prime \mu}$, and
\begin{equation} \label{eq:scalar_bound}
    |C_\text{scalar}| < \begin{cases} 0.53 ~~~ @ ~6~\text{TeV}~,~~~ 4~\text{ab}^{-1} \\
    0.26 ~~~ @ ~10~\text{TeV} ~,~~ 1~\text{ab}^{-1} \\ 0.19 ~~~ @ ~10~\text{TeV} ~,~~ 10~\text{ab}^{-1} \end{cases} ~,
\end{equation}
for the scalar Wilson coefficients $C_\text{scalar} = C_S^\mu$, $C_P^\mu$, $C_S^{\prime \mu}$, or $C_P^{\prime \mu}$. Here we give the values for the Wilson coefficients at a renormalization scale that corresponds to the center of mass energy of the collider $\mu = \sqrt{s}$.

Already with $1$~ab$^{-1}$ at a center of mass energy of 10~TeV, the constraint would be approximately as strong as the current one from $R_K$ and $R_{K^*}$~\eqref{eq:C9C10_muon_specific}. 

The constraint on the Wilson coefficients can also be translated into a sensitivity to a high new physics scale.
Assuming $\mathcal O(1)$ flavor violating new physics couplings, one has for each Wilson coefficient $C$
\begin{equation}
   \Lambda^C_\text{NP} = \left(\frac{4G_F}{\sqrt{2}} |V_{tb} V_{ts}^*| \frac{\alpha}{4\pi} |C| \right)^{-\frac{1}{2}} ~, 
\end{equation}
such that
\begin{eqnarray}
\Lambda^\text{vector}_\text{NP} &>& \begin{cases} 53~\text{TeV} ~~~ @ ~6~\text{TeV} ~,~~~ 4~\text{ab}^{-1} \\
76~\text{TeV} ~~~ @ ~10~\text{TeV} ~,~~ 1~\text{ab}^{-1} \\ 86~\text{TeV} ~~~ @ ~10~\text{TeV} ~,~~ 10~\text{ab}^{-1} \end{cases} ,  \\
\Lambda^\text{scalar}_\text{NP} &>& \begin{cases} 49~\text{TeV} ~~~ @ ~6~\text{TeV} ~,~~~ 4~\text{ab}^{-1} \\
70~\text{TeV} ~~~ @ ~10~\text{TeV} ~,~~ 1~\text{ab}^{-1} \\ 82~\text{TeV} ~~~ @ ~10~\text{TeV} ~,~~ 10~\text{ab}^{-1} \end{cases} . 
\end{eqnarray}
These results show that a muon collider has indirect sensitivity to new physics scales far above its center of mass energy and also above the scale of $\Lambda_\text{NP}^{|C| = 1} \simeq 35$~TeV, which is the generic scale associated with rare $B$ decays.

\begin{figure*}[tbh]
 \centering
  \includegraphics[width=0.46\textwidth]{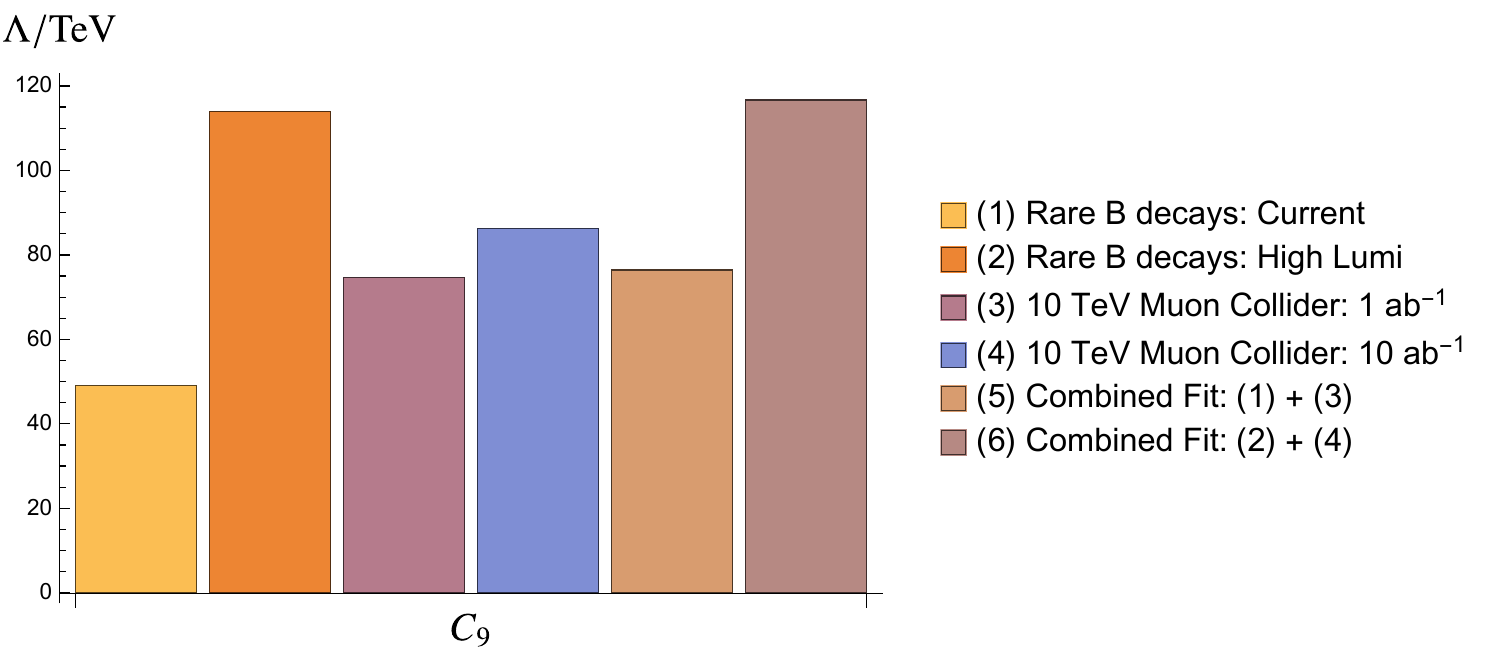} \enspace \enspace
  \includegraphics[width=0.46\textwidth]{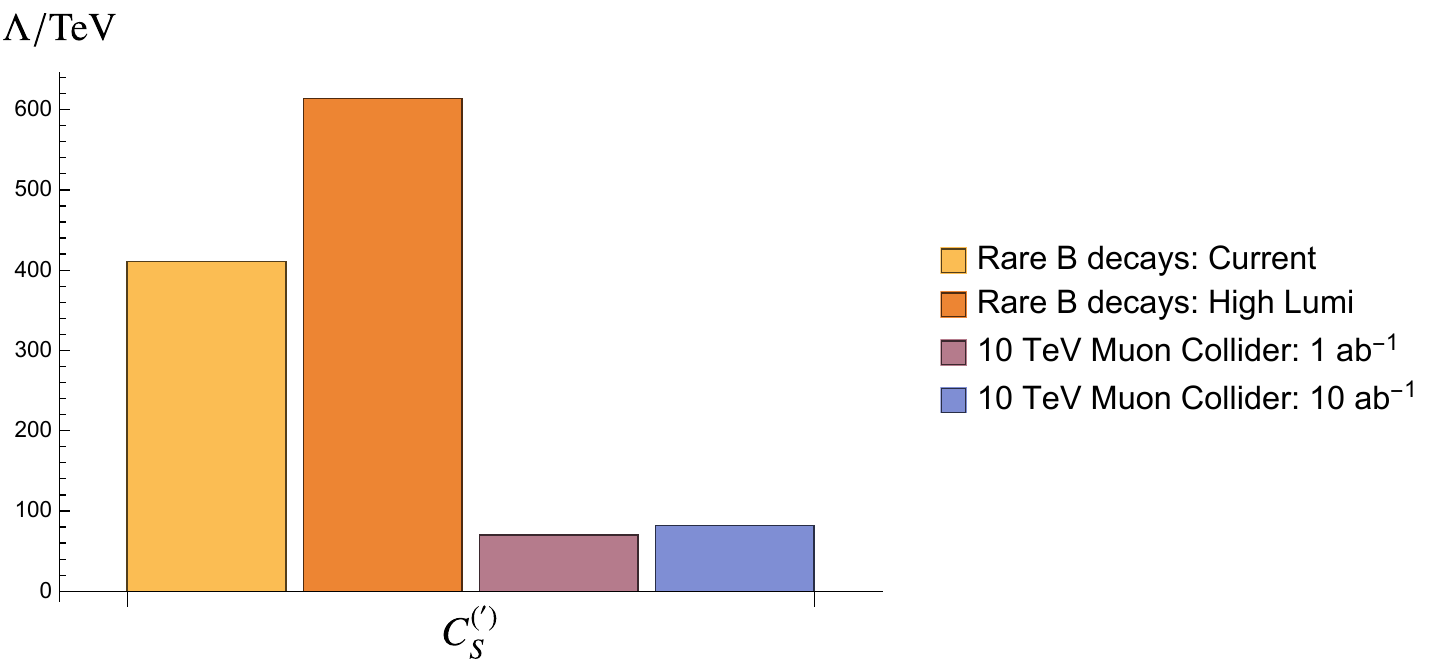}
  \caption{The new physics scales that can be probed by a muon collider and by $B$ decay data from LHCb, both current and future projections. The histogram on the left corresponds to the Wilson coefficient $\Delta C_9^\mu$, the one on the right to $C_S^{(\prime)\, \mu}$. Other vector and scalar coefficients follow identical trends.}
  \label{fig:scales_bs}
\end{figure*}

In figure~\ref{fig:scales_bs} we compare the scales that can be probed by a muon collider to those probed by current and expected future rare $B$ decay data as well as combinations thereof, for the $\Delta C_9^\mu$ coefficient in the left panel, and for $C_S^{(\prime)\, \mu}$ in the right panel.
As in \eqref{eq:vector_bound} and \eqref{eq:scalar_bound}, we assume dominance of a single Wilson coefficient.
We observe that for vector mediators the sensitivity to new physics of a 10 TeV muon collider surpasses the current LHCb sensitivity to the associated rare $B$ decays, but lags behind our projections for LHCb runs at high luminosity; vice-versa, for scalar mediators, we find that current constraints from LHCb from rare $B$ decays are anticipated to outperform a future 10 TeV collider. This is due to the fact that scalar mediators lift the helicity suppression of the $B_s \to \mu^+\mu^-$ decay, which is known to be a particularly sensitive probe of scalar new physics~\cite{Altmannshofer:2017wqy}. The two histograms to the right in the left panel illustrate that, while beneficial, combining the results of the LHC and a muon collider would only marginally strengthen the constraints from the best-performing collider.

Note that throughout our analysis, we have not made use of strange quark tagging. Therefore, in principle, the constrained cross sections do not correspond to $\sigma(\mu^+ \mu^- \to b s)$ alone, but to the combination $\sigma(\mu^+ \mu^- \to b s) + \sigma(\mu^+ \mu^- \to b d)$. Our results hold under the plausible assumption that the new physics flavor violating couplings are larger for $b\to s$ than for $b \to d$ (resembling the SM flavor hierarchies).
If that is not the case, the bounds on the Wilson coefficients in \eqref{eq:vector_bound} and \eqref{eq:scalar_bound} can be interpreted as bounds on the square sum of $b\to s$ and $b\to d$ Wilson coefficients
\begin{equation}
|C| \to \sqrt{|C^{b\to s} |^2 + |C^{b\to d} |^2} ~.
\end{equation}

It is interesting to contrast the results from sections \ref{sec:universal} and \ref{sec:muon-specific}. In the presence of a sizeable new physics signal (i.e. the scenario discussed in section~\ref{sec:universal}, with main results in figure~\ref{fig:C9_vs_C10_benchmark}), a future muon collider would be able to measure the new physics with much higher precision than LHCb. In the absence of new physics (i.e. the scenario discussed in section~\ref{sec:muon-specific}, with main results in figure~\ref{fig:scales_bs}), or for very small new physics signals, expected rare $B$ decay results from the high-luminosity runs of the LHC are more powerful in constraining new physics.
In fact, in the context of the rare $B$ decays, the new physics can interfere with the corresponding SM amplitudes, and one is linearly sensitive to small new physics. At a muon collider on the other hand, the new physics amplitude does not interfere with the backgrounds from mistags, and one is therefore only quadratically sensitive to a small new physics amplitude. 

\section{Conclusions} \label{sec:conclusions}

While the recent changes in the experimental status of the $R_{K^{(*)}}$ observables indicate lepton flavor universal physics, a few anomalies in rare $B$ decays persist. These anomalies, along with the overarching goal to search for new physics, provide a strong motivation to probe $bs\mu\mu$ interactions, while minimizing the impact of hadronic uncertainties in predicting the corresponding observable processes. With the highly complementary information that a muon collider analysis would provide, bounds on heavy new physics contributing to $b \rightarrow s \mu \mu$ decays are made far more robust in various scenarios, bolstered by the relatively clean environment of muon beams.

Here, we first reviewed the status of the global $b$ decays fit, in light of recent updates on $R_K$ and $R_{K^*}$, and identified two possible scenarios: on the one hand some new physics could be {\em lepton-flavor universal}, while addressing the $b\to s\mu\mu$ anomalies; on the other hand, some different new physics could instead be {\em muon-specific}, thus violating lepton flavor universality; in either scenario, we focused on the expected sensitivity from rare $b$ decays from the high-luminosity phase of the LHC, and then proceeded to evaluate the possible role of a future muon collider.

We computed in detail the differential cross-section for bottom-strange quark production from muon-muon collisions, including the possible effect of muon beam polarization; we then discussed and computed the irreducible Standard Model background, as well as the expected background from mis-tagged di-jet events. We then proceeded to evaluate the potential and sensitivity projections for a multi-TeV muon collider at different luminosity, center of mass energy, and beam polarization. We showed the resulting sensitivity on the plane defined by deviations of the relevant Wilson coefficients from the flavor-universal case, as well as on the potential for constraints on muon-specific new physics. Broadly, we find that a multi-TeV muon collider would be highly complementary to the LHC, and would vastly exceed the current (but not necessarily the expected high-luminosity) LHC performance in constraining new physics in $b$ decays.

\section*{Acknowledgments}

The research of WA, SAG, and SP is supported by the U.S. Department of Energy grant number DE-SC0010107.
We thank Patrick Meade and Vladimir Shiltsev for useful discussions on muon collider performance.

\begin{appendix}
\section{Renormalization Group Evolution}
\label{sec:RGEs}
For a precise sensitivity comparison of the rare $B$ decays and a muon collider, one should take into account the renormalization group running between the relevant scales.
At a muon collider, the natural scale choice for the Wilson coefficients is the center of mass energy $\mu \sim \sqrt{s}$. On the other hand, the Wilson coefficients probed by $B$ decays are typically renormalized at a low energy scale, of the order of the $b$ mass, $\mu \sim m_b$.
These scales differ by more than three orders of magnitude, and RGE running may be relevant.

We assume that the 4 fermion contact interactions in the effective Hamiltonian~\eqref{eq:effectiveHamiltonian} are the only non-zero Wilson coefficients at the scale of the muon collider. Above the electroweak scale, it is convenient to use the SMEFT operator basis from~\cite{Grzadkowski:2010es}. Our Wilson coefficients can be translated as follows~\footnote{Note that both the SMEFT coefficients $C_{\ell q}^{(1)}$ and $C_{\ell q}^{(3)}$ map onto the same combination $C_{9} - C_{10}$, and the translation into the SMEFT operators is thus not unique. Interestingly, choosing an arbitrary linear combination of $C_{\ell q}^{(1)}$ and $C_{\ell q}^{(3)}$ has no impact on the final result. We find that the difference in SMEFT RGE running of $C_{\ell q}^{(1)}$ and $C_{\ell q}^{(3)}$ is exactly compensated by threshold corrections at the electroweak scale, and at leading log accuracy, we find a unique relation between the Wilson coefficients at the collider scale and the $b$ scale.} 
\begin{eqnarray}
   [C_{\ell q}^{(1)}]_{2223} + [C_{\ell q}^{(3)}]_{2223} &=& \Delta C_{9}^\mu - \Delta C_{10}^\mu ~, \\{} 
   [C_{qe}]_{2322} &=& \Delta C_{9}^\mu + \Delta C_{10}^\mu ~, \\{} 
   [C_{\ell d}]_{2223} &=& C_{9}^{\prime \, \mu} - C_{10}^{\prime \, \mu} ~, \\{}
   [C_{ed}]_{2223} &=& C_{9}^{\prime \, \mu} + C_{10}^{\prime \, \mu} ~, \\{}
   [C_{\ell e d q}]^*_{2232} &=& 2 C_{S}^\mu ~, \\{}
   [C_{\ell e d q}]_{2223} &=& 2 C_{S}^{\prime \, \mu} ~,
\end{eqnarray}
and analogously for the Wilson coefficients with electrons and taus.

We run the Wilson coefficients from $\mu \sim \sqrt{s}$ to the electroweak scale $\mu \sim m_Z$, with the mass of the $Z$ boson $m_Z \simeq 91.2$~GeV, taking into account the impact of the gauge couplings and the top Yukawa coupling~\cite{Jenkins:2013wua, Alonso:2013hga}. After decoupling particles of electroweak mass, the Wilson coefficients are evolved further to the $b$ scale using 1-loop QED and QCD running~\cite{Jenkins:2017jig, Jenkins:2017dyc}. We find
\begin{widetext}
\begin{eqnarray}
    \Delta C_9^\mu(m_b) &\simeq& \Delta C_9^\mu(\sqrt{s}) \left[ 1 - \frac{n_\ell \alpha}{3\pi} \log\left(\frac{s}{m_b^2}\right) - \frac{\alpha}{16 \pi s_W^2} \left(\frac{1}{c_W^2} + 2 +\frac{m_t^2}{2 m_W^2} \right) \log\left(\frac{s}{m_Z^2}\right) \right] \nonumber \\
    && + \Delta C_{10}^\mu(\sqrt{s})\left[ \frac{\alpha}{2\pi} \log\left(\frac{s}{m_b^2}\right) +  \frac{\alpha}{16\pi s_W^2} \left( \frac{1}{c_W^2} + 2 \right)\left(1 - 4s_W^2\right) \log\left(\frac{s}{m_Z^2}\right) \right] ~, \\[8pt]
    \Delta C_{10}^\mu  (m_b) &\simeq& \Delta C_{10}^\mu(\sqrt{s}) \left[ 1 - \frac{\alpha}{16\pi s_W^2} \left( \frac{1}{c_W^2} + 2 + \frac{m_t^2}{2 m_W^2} \right) \log\left(\frac{s}{m_Z^2}\right) \right] \nonumber \\
    && + \Delta C_9^\mu(\sqrt{s})\left[ \frac{\alpha}{2\pi} \log\left(\frac{s}{m_b^2}\right)  + \frac{\alpha}{16\pi s_W^2} \left(\frac{1}{c_W^2} + 2\right) \left(1 - 4 s_W^2\right) \log\left(\frac{s}{m_Z^2}\right) \right] ~, \\[8pt]
    C_9^{\prime\,\mu}(m_b) &\simeq& C_9^{\prime\,\mu}(\sqrt{s}) \left[ 1 - \frac{n_\ell \alpha}{3\pi} \log\left(\frac{s}{m_b^2}\right) - \frac{\alpha}{8\pi c_W^2} \log\left(\frac{s}{m_Z^2}\right) \right] \nonumber \\
    && - C_{10}^{\prime\,\mu}(\sqrt{s})\left[ \frac{\alpha}{2\pi} \log\left(\frac{s}{m_b^2}\right) - \frac{\alpha}{8\pi c_W^2} \left( 1 - 4s_W^2 \right) \log\left(\frac{s}{m_Z^2}\right) \right] ~, \\[8pt]
    C_{10}^{\prime\,\mu}(m_b) &\simeq& C_{10}^{\prime\,\mu}(\sqrt{s}) \left[ 1 - \frac{\alpha}{8\pi c_W^2} \log\left(\frac{s}{m_Z^2}\right) \right]  \nonumber \\
    && 
    - C_9^{\prime\,\mu}(\sqrt{s})\left[ \frac{\alpha}{2\pi} \log\left(\frac{s}{m_b^2}\right) - \frac{\alpha}{8\pi c_W^2} \left(1 - 4 s_W^2 \right) \log\left(\frac{s}{m_Z^2}\right) \right] ~, \\[8pt]
    C_S^\mu(m_b) &\simeq& C_S^\mu(\sqrt{s}) \left(\frac{\alpha_s(m_b)}{\alpha_s(m_Z)}\right)^\frac{12}{23} \left(\frac{\alpha_s(m_Z)}{\alpha_s(\sqrt{s})}\right)^\frac{4}{7} \left[ 1 + \frac{5\alpha}{6\pi} \log \left(\frac{s}{m_b^2}\right) + \frac{\alpha}{3\pi} \left(\frac{1}{c_W^2}-\frac{5}{2}\right) \log\left(\frac{s}{m_Z^2}\right) \right]~, \\[8pt]
    C_S^{\prime\,\mu}(m_b) &\simeq& C_S^{\prime\,\mu}(\sqrt{s}) \left(\frac{\alpha_s(m_b)}{\alpha_s(m_Z)}\right)^\frac{12}{23} \left(\frac{\alpha_s(m_Z)}{\alpha_s(\sqrt{s})}\right)^\frac{4}{7} \left[ 1 + \frac{5\alpha}{6\pi} \log \left(\frac{s}{m_b^2}\right) + \frac{\alpha}{3\pi} \left(\frac{1}{c_W^2} - \frac{5}{2} - \frac{3m_t^2}{32s_W^2 m_W^2}\right) \log\left(\frac{s}{m_Z^2}\right) \right] ~.
\end{eqnarray}
\end{widetext}
Where $n_\ell = 1$ in the scenario with muon-specific Wilson coefficients and $n_\ell = 3$ in the scenario with lepton universal coefficients. 
In these expressions, $s_W$ and $c_W$ are the sine and cosine of the weak mixing angle, $\alpha$ is the fine structure constant, and $\alpha_s$ is the strong coupling constant. 
Note that only the scalar coefficients $C_S = -C_P$ and $C_S^\prime = C_P^\prime$ experience QCD running. In the above expressions, we re-summed the QCD logarithms but used a leading logarithmic approximation for the (much smaller) electroweak RGE corrections.

In the lepton universal scenario, running to the low scale preserves universality, and the expressions above also hold analogously for the electron and tau coefficients. In the muon-specific scenario, the RGE running induces operators with electrons and taus. This is, in principle, relevant in the context of the rare $B$ decays, as the LFU ratios $R_{K^{(*)}}$ are to a good approximation sensitive to the differences of the muon and electron Wilson coefficients $C_i^\mu - C_i^e$.
In this case, the non-zero electron coefficients at the $b$ scale are
\begin{eqnarray}
    \Delta C_9^e(m_b) &\simeq& - \Delta C_9^\mu(\sqrt{s}) \frac{\alpha}{3\pi} \log\left(\frac{s}{m_b^2}\right) ~, \\
    C_9^{\prime \, e}(m_b) &\simeq& - C_9^{\prime \, \mu}(\sqrt{s}) \frac{\alpha}{3\pi} \log\left(\frac{s}{m_b^2}\right) ~.
\end{eqnarray}

In practice, we find that the RGE running has a small impact. The shifts between the Wilson coefficients $C_{9,10}^{(\prime)}$ at the low scale $\mu \simeq m_b \simeq 4.2$~GeV and the high scale $\mu \simeq \sqrt{s} = 10$~TeV are typically around 5-10\%. We do include the running of $C_{9,10}^{(\prime)}$ in the leading log approximation in our numerical analysis for completeness.
In the case of the scalar coefficients $C_{S,P}^{(\prime)}$, we take into account the small electroweak running and the QCD running, which is an $\mathcal O(1)$ effect. We note that tools that can perform the running numerically are, in principle, available~\cite{Aebischer:2018bkb, Fuentes-Martin:2020zaz, DiNoi:2022ejg}.

\section{Updated Rare $B$ Decay Fit}
\label{sec:GlobalFit}

In this appendix, we briefly describe our rare $B$ decay fit that incorporates the recent LHCb results of $R_K$ and $R_{K^*}$~\cite{LHCb:2022qnv, LHCb:2022zom} (see also~\cite{Ciuchini:2022wbq, Greljo:2022jac, Allanach:2022iod, Alguero:2023jeh, Wen:2023pfq, Allanach:2023uxz} for other recent discussions). 

We perform our own global fit using \verb|flavio|~\cite{Straub:2018kue} (version 2.3.3) and include the following set of experimental results
\begin{itemize}
\item branching ratio measurements of $B_s \to \mu^+ \mu^-$ from CDF~\cite{CDF:2011wlz}, ATLAS~\cite{ATLAS:2018cur}, and LHCb~\cite{LHCb:2021vsc};
\item branching ratios measurements of $B\to K\mu^+\mu^-$, $B\to K^*\mu^+\mu^-$, and $B_s\to \phi \mu^+\mu^-$ from CDF~\cite{Miyake:2012exl}, Belle~\cite{BELLE:2019xld}, CMS~\cite{CMS:2015bcy}, and LHCb~\cite{LHCb:2014cxe, LHCb:2016ykl, LHCb:2021zwz};
\item angular observables of the decays $B\to K^*\mu^+\mu^-$ and $B_s\to \phi \mu^+\mu^-$ from CDF~\cite{Miyake:2012exl}, ATLAS~\cite{ATLAS:2018gqc}, CMS~\cite{CMS:2017ivg}, and LHCb~\cite{LHCb:2020lmf, LHCb:2020gog, LHCb:2021xxq};
\item branching ratios and angular observables of $\Lambda_b \to \Lambda \mu^+ \mu^-$ from CDF~\cite{Miyake:2012exl} and LHCb~\cite{LHCb:2015tgy, LHCb:2018jna};
\item lepton flavor universality tests in rare B meson decays from Belle~\cite{Belle:2016fev, BELLE:2019xld} and LHCb~\cite{LHCb:2021lvy}.
\end{itemize}
To reduce hadronic uncertainties on the semi-leptonic branching ratios and angular observables, we only take into account $q^2$ bins below 6~GeV$^2$, as well as broad bins above the narrow charmonium resonances that span the entire available kinematic range. To be conservative, we do not use \verb|flavio|'s default values for the CKM matrix elements, but instead use the PDG values~\cite{ParticleDataGroup:2022pth} $|V_{cb}| = (40.8 \pm 1.4)\times 10^{-3}$ and $|V_{ub}| = (3.82 \pm 0.20)\times 10^{-3}$, which are a conservative average of inclusive and exclusive determinations with inflated uncertainties. 

In addition to the observables listed above, we implement the recent CMS results on the $B_s \to \mu^+ \mu^-$ decay from~\cite{CMS:2022dbz}, as well as the latest LHCb results on $R_K$ and $R_{K^*}$~\cite{LHCb:2022qnv, LHCb:2022zom} which were not yet included in version 2.3.3 of \verb|flavio| (but have been added in the latest update). These results have a significant impact on the fit. In figure~\ref{fig:global_fit} in the main text, we show the result of our fit in the standard plane of muon specific $C_9$ and $C_{10}$ Wilson coefficients. 

\end{appendix}

\bibliography{main}{}

\end{document}